\documentstyle[12pt,epsf]{article}
\begin{document}
\footskip50pt
\parindent=12pt
\parskip=.4cm
\input amssym.def
\input amssym
\def\tC{\tilde{C}}
\def\bR{{\Bbb R}}
\def\a{\"{a}}
\def\o{\"{o}}
\def\u{\"{u}}
\def\eps{\varepsilon}
\def\l{\langle}
\def\r{\rangle}
\def\g{g^{\mu \nu}}
\def\d{{\rm d}}
\def\D{{\cal D}}
\def\L{{\cal L}}
\def\S{{\cal S}}
\def\F{{\cal F}}
\def\T{{\cal T}}
\def\X{{\cal X}}
\def\Y{{\cal Y}}
\def\ome{{\cal A}}
\def\Ga{\Gamma}
\def\pr{\partial}
\def\O{{\cal O}}
\def\R{{\cal R}}
\def\nab{\nabla}
\def\I{{\cal I}}
\def\J{{\cal J}}
\def\E{{\cal E}}
\def\hU{\hat{U}}
\def\hV{\hat{V}}
\def\hh{{1\over 2}}
\def\half{{\textstyle{ 1 \over 2}}}
\def\quar{{\textstyle{ 1 \over 4}}}
\def\ts{\textstyle}
\def\A{{\cal A}}
\def\B{{\cal B}}
\def\C{{\cal C}}
\def\de{\delta}
\def\si{\sigma}
\def\ga{\gamma}
\def\la{\lambda}
\def\ka{\kappa}
\def\be{\beta}
\def\topcirc{\mathaccent"7017}
\def\oT{{\topcirc T}}
\def\oG{{\topcirc G}}
\def\oD{{\topcirc \D}}
\def\oDel{{\topcirc \Delta}}
\renewcommand{\theequation}{\thesection.\arabic{equation}} 

\begin{flushright}
DAMTP/96-108, hep-th/9704108
\end{flushright}

\begin{center}
{\Large \bf Conformally Covariant Differential Operators:}\break
{\Large \bf Properties and Applications} 
\end{center}

\vspace{1.5cm}

\begin{center}
Johanna Erdmenger\footnote{new permanent 
address: Institut f\u r theoretische Physik, 
Universit\a t Leipzig, Augustusplatz 10/11, D - 04109 Leipzig, Germany. \
e-mail: erd@tph204.physik.uni-leipzig.de }

Department of Applied Mathematics and Theoretical Physics, University
of Cambridge, Silver Street, Cambridge CB3 9EW, England. \ e-mail: 
je10001@damtp.cam.ac.uk  

\end{center}

\vspace{1.5cm}

\begin{center} \bf Abstract \end{center}

\rm \normalsize

We discuss
conformally covariant differential operators, which under local
rescalings of the metric, $\de_\si g^{\mu\nu} = 2 \si g^{\mu\nu}$, 
transform according to
$\de_\si \Delta
= r \Delta \si + (s-r)\si \Delta$ for some $r$ if $\Delta$ is $s\,$th order. 
It is shown that the flat space restrictions of 
their associated Green functions have forms
which are strongly constrained by flat space conformal invariance. 
The same applies to the variation of the Green functions with respect to the 
metric. The general
results are illustrated by finding the flat space Green function and also
its first variation for previously found second order conformal differential 
operators acting on $k$-forms in general dimensions. Furthermore we construct 
a new second order
conformally covariant operator acting on rank four tensors with the symmetries
of the Weyl tensor whose Green function is similarly discussed. We also
consider fourth order operators, in particular a fourth order operator
acting on scalars in arbitrary dimension, which has a Green function with the
expected properties.
The results obtained here for conformally covariant diffe\-rential 
operators are
generalisations of standard results for the two dimensional Laplacian on
curved space and its associated Green function which is used in the Polyakov
effective gravitational action. It is hoped that they may have similar
applications in higher dimensions.
 
\vspace{1ex}

\begin{flushleft}
\newline Quantum field theory.
\end{flushleft}

\newpage

\section{Introduction}
 
\setcounter{equation}{0}

In discussions of the effects of quantum matter on
gravity the effective action $W[g]$, for some fixed background metric
$g_{\mu\nu}$, is of crucial interest. 
It is a scalar under diffeomorphisms, i.e.~local coordinate
reparametrisations when of course $g_{\mu\nu}$ transforms as a tensor field. 
The energy momentum tensor is defined by the response of
the field theory to variations of the metric so that its expectation value
for the background metric $g_{\mu\nu}$ is given in terms of the effective 
action by
\begin{equation} \label{tdw}
\l T_{\mu \nu} (x) \r  =  - \frac{2} {\ts{\sqrt{g(x)}}} \frac{\delta}{\delta
g^{\mu \nu} (x) } W[g] \, .
\end{equation}
Here we consider quantum field theories where the trace of the energy
momentum tensor has zero contributions from the quantum matter fields.
Such field theories on curved space are expected to be also Weyl invariant,
i.e. invariant under local rescalings of the metric.
However in even dimensions there are local anomalies such that the
effective action is not invariant under Weyl rescalings and the energy
momentum tensor expectation value acquires an anomalous trace involving
the curvature.
In two dimensions this anomalous trace is simply given by
\begin{equation} \label{ano2}
g^{\mu\nu}\l T_{\mu \nu} \r = \frac{c}{24 \pi} \, R \, ,
\end{equation}
with $R$ the Ricci scalar and $c$  the Virasoro central charge. 
In four dimensions the expression for the anomalous trace is more complicated.
It contains at least two linearly independent dimension 4 scalars $F,G$ formed
from the metric so that
\begin{equation} \label{FG}
g^{\mu \nu} \l T_{\mu \nu}\r = 
- \beta_a \, F - \beta_b \, G  
\end{equation}
where $\beta_a$ and $\beta_b$ are coefficients depending on the particular
theory and
\begin {eqnarray}
 F & = & 
 R^{\alpha \beta \gamma \delta} R_{ \alpha \beta
\gamma \delta} - \frac{4}{d-2} \, R^{\alpha \beta} R_{\alpha \beta} +
\frac{2}{(d-2)(d-1)} \, R^2 \; = \; C^{\alpha \beta \gamma \delta} C_{\alpha
\beta \gamma \delta} \, \mbox{,} \label{F}\\
G & = &  R^{\alpha \beta \gamma \delta} R_{ \alpha \beta
\gamma \delta} - 4 \, R^{\alpha \beta} R_{\alpha \beta} + R^2 
\; = \; \textstyle{1 \over 4} \eps^{\mu \nu \sigma \rho } \eps_{\alpha \beta
\gamma \delta} R^{\alpha \beta}{}_{\, \mu \nu}
R^{\gamma \delta}{}_{\, \sigma \rho} \label{GB}
\, \mbox{.} 
\end{eqnarray}
$C_{\alpha \beta \gamma \delta}$ is the Weyl tensor which is given by
\begin{equation}
C_{\alpha \beta \gamma \delta} = R_{\alpha \beta \gamma \delta} 
- {2} \left( g_{\alpha [ \gamma} K_{\delta ] \beta} - 
g_{\beta [ \gamma} K_{\delta ] \alpha} \right)  \, , \label{Weylg}
\end{equation}
with
\begin{equation}
K_{\alpha \beta} = \frac{1}{d-2} \, \bigg( R_{\alpha \beta} -
\frac{1}{2(d-1)} g_{\alpha \beta} R \bigg) \, .
\end{equation}
$F$ and $G$ are the only necessary gravitational contributions to the trace 
anomaly in general. 
Although a term proportional to $R^2$ might be expected in the trace
(\ref{FG}), it must be absent here as a consequence of Wess-Zumino consistency
conditions \cite{WZ}.
Moreover possible contributions to the trace anomaly proportional to $\nabla^2
 R$  may be cancelled by adding a local term to the effective action.

In two dimensions there is a well-known unique 
expression for the effective action due to Polyakov
\cite{Polyakov}, which is obtained by integrating the
anomaly (\ref{ano2}),
\begin{eqnarray} \label{polyakov}
W^P[g] = {} - {c\over 96\pi} \int \! \! \int \! \d^2 x \d^2 y \,
{\ts\sqrt{g(x)}}R(x) G (x,y) {\ts\sqrt{g(y)}}R(y) \, , 
\end{eqnarray}
with
\begin{eqnarray}
- {\ts\sqrt {g(x)}} \nabla^2{}_{\!\!\! x} G (x,y) = \de^2(x-y) \, .
\label{2op}
\end{eqnarray}
The Polyakov  action is manifestly a diffeomorphism invariant scalar. 
It has a
non-local structure as it involves the Green function of a second order
differential operator $\nabla^2$ which transforms simply under local rescalings
of the metric $ \delta_\si g^{\mu \nu} = 2 \si g^{\mu \nu}$,
\begin{equation}
\delta_\sigma (\sqrt g \nabla^2) = 0 \, .
\end{equation}
The Green function $G(x,y)$ associated to $\nabla^2$
is therefore invariant under local rescalings of the metric,
\begin{equation}
 \delta_{\si} G (x,y) = 0 \, .\; 
\end{equation}
With $\de_\si \sqrt g
R  = 2\sqrt g \nabla^2 \si$ it is easy to check that the variation of the 
Polyakov action (\ref{polyakov}) gives the conformal anomaly (\ref{ano2}).

In four dimensions 
it is also natural to consider possible non-local constructions for the 
effective action involving Green functions $G_\Delta$ associated with
differential operators $\Delta$ with simple properties under
Weyl rescalings and which reproduces the anomaly (\ref{FG}).
To this end  in this paper we analyse such differential operators, which
we refer to here as conformally covariant differential operators, and
derive some general properties of their associated Green functions.
The simplest such operator in $d$ dimensions is the second order operator
\begin{equation} \label{2dop}
\Delta_{2} = - \nabla^2 + {d-2 \over 4(d-1)}R
\end{equation}
acting on scalar fields, with the variation
\begin{equation}
\delta_\si \Delta_{2} =   \half (d+2) \sigma \Delta_{2} - \half (d-2)
\Delta_{2} \sigma    \, .  
\end{equation}
Its associated Green function satisfies
\begin{equation}
\de_\si G_2(x,y) = \half (d-2) ( \si(x) + \si(y) ) G_2(x,y) \, .
\label{cvarG2}
\end{equation}
Clearly in two dimensions $\Delta$ reduces to $-\nabla^2$ and the Green 
function  to $G(x,y)$ which featured in the above result for the
Polyakov action.
We investigate here generalisations to
operators acting on various tensor fields in arbitrary dimensions $d$.
The construction of conformally  covariant differential operators, and also
conformal invariants
involving the Riemann curvature tensor, have been discussed extensively
by mathematicians. In particular we may mention the results, in the 
general framework of differential geometry, found by  Branson \cite{Branson},
Fefferman and Graham \cite{Fefferman}, Parker and Rosenberg
\cite{Rosenberg}, and by W\u nsch \cite{Wuensch}. 
Here we use especially the second order differential  operator
acting on $k$-forms in $d$ dimensions first obtained by Branson \cite{Branson}. 
We also construct another conformally covariant
second order differential operator which
acts on tensor fields with the symmetries of the Weyl tensor defined in
(\ref{Weylg}).

A crucial observation for our subsequent
discussion is that theories defined on curved
space which are invariant under diffeomorphisms and also Weyl rescalings of
the metric are expected to be conformally invariant when reduced to flat 
space, when $g_{\mu\nu} = \delta_{\mu\nu}$. 
Conformal invariance is a strong symmetry constraint for field
theories which allows for exact results for the two and three point correlation
functions. Besides massless free field theories the physical relevance 
of conformal invariance is given by the fact that it should be
realised for interacting quantum
field theories at renormalisation group fixed points.

This property is illustrated by 
the Polyakov action  since it is in agreement with  conformal 
flat Euclidean space correlation
functions  \cite{EO}. We define the two point 
function on flat space to be
\begin{equation}\label{tdw2}
\l T_{\mu\nu} (x) T_{\sigma\rho}(y) \r \, = \, 
4 \frac{ \delta^2}{\delta g^{\mu \nu} (x) \delta g^{\si \rho}(y)} \, W^P[g]
\bigg|_{g= \delta} \, ,
\end{equation} 
with a similar expression for the three point function.
Although $W^P[g]$ itself is zero on flat space where the curvature
vanishes, second or higher functional derivatives of $W^P[g]$ are non-zero
even on flat space.  
Using 
$\delta_{g} (\sqrt{g} R) = - \sqrt{g} ( \nabla_{\mu} \nabla_{\nu} -
g_{\mu \nu} \nabla^2)\delta\g $ and  restricting to flat space, when
$G (x,y)|_{g=\delta} = - {\rm ln} \mu^2 (x-y)^2/4 \pi$
with $\mu$ an arbitrary scale, we obtain
\begin{eqnarray}
\l T_{\mu\nu} (x) T_{\sigma\rho}(y) \r \,
& = & -{c\over 48\pi^2}S^x 
{}_{\!\! \mu\nu}
S^y {}_{\!\! \sigma \rho} {\rm ln} (x-y)^2 \, , \; \; S_{\mu \nu} =
\partial_{\mu} \partial_{\nu} - \delta_{\mu \nu} \partial^2 \, .
\end{eqnarray}
With conventional complex coordinates $z$ on the plane, so that 
$x^2=z{\bar z}$, and defining
$T(z) = -2 \pi T_{zz} (x)$, this gives
rise to the standard two dimensional conformal field theory result
\begin{equation} \label{2d2}
\l T(z_1) T(z_2) \r = - \frac{c}{12}\, 
\pr_{z_1}{}^{\!\!\! 2} \pr_{z_2}{}^{\!\!\! 2}
\ln (x_1-x_2)^2 = \frac{\half c}{(z_1-z_2)^4} \, . 
\end{equation}
Thus this calculation shows that the conformal energy momentum tensor 
two point function on flat space
is completely determined by the conformal anomaly (\ref{ano2}) on curved space.
If we use for the variation of the Green function
\begin{equation} 
\frac{\de}{\de g^{zz} (x_3)} G(x_1,x_2) \bigg |_{g=\de} = - \frac{1}{(4\pi)^2}
\, \frac{1}{(z_1-z_3)(z_2-z_3)} \, ,
\label{varG}
\end{equation}
then the standard result for the three point function at non coincident points,
\begin{eqnarray} \!\!\!\!\!
\l T(z_1) T(z_2) T(z_3) \r\!\!\! &=&\!\!\!  - \frac{c}{3} \bigg \{
{1\over (z_1-z_3)^3 (z_2 -z_3)^3}\! + \!{1\over (z_2-z_1)^3 (z_3 -z_1)^3}
\!+\! {1\over (z_1-z_2)^3 (z_3 -z_2)^3} \bigg \}
\nonumber \\
&=& \!\!\!\frac{c}{(z_1-z_2)^2 (z_2-z_3)^2
(z_3-z_1)^2 } \, ,
\end{eqnarray}
may be similarly obtained 
by varying the Polyakov action (\ref{polyakov}) three times with respect
to the metric \cite{EO}. 

In four dimensions  there is no analogue of the Polyakov action at present
which has all its properties, i.e.~which just yields the conformal anomaly
(\ref{FG}) 
under Weyl rescalings, and which also 
leads to conformally invariant correlation
functions for the energy momentum tensor on flat space. 
An action constructed by Riegert \cite{Riegert}, which involves the Green
function of a conformally covariant fourth order operator acting on scalars, 
generates the conformal anomaly upon Weyl rescalings. However this does not 
lead to
conformal correlation functions on flat space. As discussed in \cite{EO}, 
this is due to the insufficiently rapid fall off at long-distances 
of the Riegert
action, which is responsible for surface terms spoiling the conformal
invariance Ward identities. This is related to the fact, pointed out by Deser
\cite{Deser2}, that the Green function for the fourth order
differential operator which appears in the  Riegert action has a double pole.
A general construction for the four dimensional effective action to third
order in the curvature, in terms of a 
basis for non-local invariants, has been constructed by Barvinsky et
al. \cite{Barvinsky}. This is unfortunately very complicated
and it is not at all clear which  minimal  linear
combination of these terms is sufficient for an action with the
required properties.

The organisation of this paper is as follows: We begin by discussing conformal
diffe\-rential operators on curved space and their associated Green functions 
in all generality in section 2. It is shown how the form of the variation
of the Green function with respect to the metric is constrained by
conformal invariance when reduced to flat space.
In section 3 we apply our general results to
the second order differential operator on $k$-forms constructed by Branson. 
In section 4 we discuss some previous results for scalars which transform
homogeneously under local rescalings of the metric, i.e. infinitesimally
$\de_\si \O= \eta \si \O$ for dimension $\eta$. We rederive the expression
for a scalar of dimension 6 in which was found by Fefferman and Graham
\cite{Fefferman} in  addition to the two dimension 6 scalars
which may be constructed trivially in terms of the Weyl tensor alone. 
With similar results we then construct a conformal second order differential 
operator acting on tensors
with Weyl symmetry in section 5. We also discuss the Green function of this
operator. In section 6 we consider fourth order operators, in particular
an operator acting on scalars which reduces to the operator whose Green
function is used in the Riegert action in four dimensions. Results compatible
with conformal invariance are again obtained. Finally in section 7 we
describe how the Riegert action is constructed but also that the fourth
order operator has 4 as a critical dimension so that conformal invariant
results are not in general obtained when reducing to flat space.

\section{Conformally Covariant Differential Operators}

\setcounter{equation}{0}

We consider fields $\O^i(x)$ which are sections of tensor bundles over
$d$-dimensional Riemannian space with metric $g^{\mu \nu}$. Under
diffeomorphisms $ \de x^\mu = v^\mu(x) $, 
i.e.~local coordinate reparametrisations, these
fields transform as
\begin{equation}
\de_v \O^i(x) = \L_{v(x)} \O^i(x) \, , \; \; \; \;
\end{equation} 
with $\L_v$ the appropriate 
Lie derivative. Acting on scalars $\varphi$ and on the
metric $g^{\mu \nu}$, $\L_v$ is given by
\begin{eqnarray}
\L_v \varphi = v^\lambda \pr_\lambda \varphi \, , \; \; \; \; 
\L_v g^{\mu \nu} &=& v^\lambda \pr_\lambda g^{\mu \nu} - \pr_\lambda v^\mu
g^{\lambda \nu}- \pr_\lambda v^\nu g^{\mu \lambda}
\,  \nonumber\\ &=& - \nabla^\mu v^\nu - \nabla^\nu
v^\mu \, . \label{lie}
\end{eqnarray}

It is also convenient to define 
the field $\bar{\O}_i(x)$ conjugate to $\O^i(x)$ 
such that $ \bar{\O}_i\O^i$ is a scalar,
$\de_v(  \bar{\O}_i\O^i) = v^\mu \pr_\mu ( 
 \bar{\O}_i\O^i )$. The transformation property is written as
\begin{equation}
\de_v \bar{\O}_i(x) = \bar{\L}_{v(x)} \bar{\O}_i(x) \; .
\end{equation}
Thus corresponding to (\ref{lie})
\begin{equation} \bar{\L}_v \varphi = {\L}_v \varphi \, , \; \; \;
\bar{\L}_{v} g_{\mu \nu} = 
v^\lambda \pr_\lambda g_{\mu \nu} + \pr_\mu v^\lambda g_{\lambda \nu}
+ \pr_\nu v^{\lambda} g_{\mu \lambda} \, 
\end{equation}
when acting on scalars or on the metric.
Furthermore under local Weyl rescalings of the metric,
\begin{equation}
\de_\si g^{\mu \nu}(x) = 2 \, \si(x) g^{\mu \nu} (x) \, ,
\end{equation}
we require
 \begin{equation} \label{ow}
\de_{\si   } \O^i(x) = - r \, \si(x) \O^i(x) \, , \; \; \; \;
\de_{\si   } \bar{\O}_i(x) = - \bar{r} \, \si(x) \bar{\O}_i(x) \, ,
\end{equation}
with some real numbers $r, \bar{r}$.

Now let ${\Delta}$ be an elliptic differential operator of
order $s$. Under diffeomorphisms and Weyl rescalings we then assume
\begin{eqnarray}
\de_v({\Delta} \O)^i(x) &=&  \L_{v(x)} (\Delta \O)^i(x) \\
\de_\si({\Delta} \O)^i(x) &=&  (s-r) \si(x) 
({\Delta} \O)^i(x) \, . \label{dw}
\end{eqnarray}
(\ref{ow}) and (\ref{dw}) imply
\begin{equation} \label{sd}
\de_\si {\Delta} = r \Delta \si + (s-r) \si \Delta \, .
\end{equation}

Since $\de_v \sqrt{g} = \sqrt g \nabla {\cdot v}$
the expression
\begin{equation}
S_\O = \int \! \d^d x \, \sqrt g \,  \bar{\O}_i (\Delta \O)^i 
\end{equation}
is then an invariant scalar, $\de_v S_\O = 0$, and moreover this is also Weyl
invariant, $\de_\si S_\O = 0$, if
\begin{equation}
\bar{r} + r - s = \, -d \, .
\end{equation}

The equation
\begin{equation} \label{dgreen}
{\ts\sqrt {g(x)}} (\Delta_x G_\Delta)^i{}_j (x,y) = \delta^i{}_j \delta^d(x-y)
\end{equation}
defines the Green function $G_\Delta{}^i{}_j(x,y)$ of the operator $\Delta$. 
$\de^i{}_j$ is the identity for the space of tensors under consideration.
Under diffeomorphisms this Green function transforms as
\begin{equation} \label{dgreen2}
\de_v G_\Delta{}^i{}_j(x,y) = \L_{v(x)} G_\Delta{}^i{}_j(x,y) +
\bar{\L}_{v(y)} G_\Delta{}^i{}_j(x,y) \, . 
\end{equation}
The right hand side of (\ref{dgreen}) is invariant under Weyl
rescalings, which implies together with (\ref{sd}) that
\begin{equation} \label{sg}
\de_\si G_\Delta{}^i{}_j(x,y) = (d-r) \si(x) G_\Delta{}^i{}_j(x,y) -
(s-r) \si(y) G_\Delta{}^i{}_j(x,y)  \, .
\end{equation}

Thus the conditions for Weyl and diffeomorphism invariance of the Green 
function implied by
(\ref{dgreen}), (\ref{dgreen2}) and (\ref{sg}) are 
\begin{eqnarray} \label{cinv1}
  \Big( (d-r) \si(x)  - (s-r) 
\si(y) \Big) G_\Delta{}^i{}_j(x,y)  \hspace{5cm}
& & \nonumber\\ \hspace{2cm} {} + 2 \int \! \d^d z \, \si(z)
g^{\alpha \be}(z) \frac{\de}{\de g^{\alpha \be}(z)}  G_\Delta{}^i{}_j(x,y)
&=& 0 \, , \\ \label{cinv2}
( \L_{v(x)} + \bar{\L}_{v(y)} ) 
G_\Delta{}^i{}_j(x,y) \, 
+  \, \int \!\d^d z \, \L_{v(z)} g^{\alpha \beta}(z)
\frac{\de}{\de g^{\alpha \beta}(z)} G_\Delta{}^i{}_j(x,y)
& =&  0 \, .
\end{eqnarray}
The sum of these two equations gives a non-trivial condition on
$G_\Delta$ for a fixed metric  if we restrict $v, \, \si = \si_v$ by
\begin{equation} \label{ccond}
\L_v g^{\alpha \beta} + 2 \si_v g^{\alpha \beta} =0 \, , \;\;\; \si_v =
\nabla {\cdot v}/d \,  .
\end{equation}
With this condition, (\ref{cinv1}) and (\ref{cinv2}) yield 
\begin{equation} \label{gconf}
\Big(\L_{v(x)} + \bar{\L}_{v(y)} +(d-r) \si_v(x) - (s-r)
\si_v(y) \Big) G_\Delta{}^i{}_j(x,y) = 0 \, ,
\end{equation}
although $v, \, \si_v$ must be constrained so that surface terms can be 
neglected.
Restricting to flat Euclidean space, (\ref{ccond})
gives the conformal Killing equation
\begin{equation} 
\pr_\mu v_\nu + \pr_\nu v_\mu = 2 \si_v \delta_{\mu \nu} \, ,
\end{equation} 
which ensures that $\de x_\mu = v_\mu(x)$ is a conformal transformation, and
any Green function satis\-fying (\ref{gconf}) is conformally
covariant on flat space.

On flat space by translational invariance the
Green function $G_\Delta(x,y)$ depends only on $x-y$, so that denoting the
flat space Green function by $\oG_\Delta$ we have
\begin{equation}
\oG_\Delta{}^{i}{}_{j}(x-y) = G_\Delta{}^{i}{}_{j}(x,y) \Big|_{g=\de} \, .
\end{equation}
However applying the results for conformal two point functions
developed in \cite{OP} gives an explicit form for $\oG_\Delta$ for conformal
differential operators,
\begin{equation} \label{ggreen}
\oG_\Delta{}^{i}{}_{j}(x) = C_\Delta \, \frac{D^i{}_j
\Big(I(x)\Big)} {(x^2)^{{1 \over 2} (d-s)}} \, ,
\end{equation}
where ${D^i{}_j(I(x))}$ is here the appropriate representation of the inversion
tensor acting on the fields $\O^i$ and $C_\Delta$ is some constant coefficient
depending on $d$ and the particular tensor representation. 
Inversions are conformal coordinate transformations for which
\begin{equation}
x'{}_{\!\mu} = \frac{x_{\mu}}{ x^2} \; \mbox{,} \;
\end{equation}
and the fundamental representation of the inversion tensor is given by
\begin{equation}
x'{}_{\!\mu} = x^2 I_{\mu \nu}(x) x_\nu \, , \; \; \;
 I_{\mu \nu} (x) = \delta_{\mu \nu} - 
\frac{2 x_{\mu} x_{\nu}}{x^2} \, ,\; \; {\rm det} I = - 1 \, .
\end{equation}
The inversion is a conformal transformation not connected to the
identity. Its significance for the two point functions is due to the
fact that it plays the role of parallel transport for conformal
transformations.

The variation of the Green function $G_\Delta$ with respect to a
change in the metric is given by
\begin{equation} \label{varg}
\de_g 
G_\Delta{}^i{}_j(x,y) = - \int \! \d^d z \, G_\Delta{}^i{}_k(x,z)
\Big( {\de_g (\sqrt g \Delta_z) G_\Delta } \Big){}^k{}_j (z,y)
\, .
\end{equation}
If we thereby define
\begin{equation} 
G'{}_{\!\Delta}{}^i{}_{j,\alpha\beta}(x,y;z) = 
\frac{\de}{\de g^{\alpha \be} (z)} G_\Delta{}^{i}{}_{j}(x,y) \, ,
\label{Gvar1}
\end{equation}
then scale and diffeomorphism invariance imposed by virtue of (\ref{cinv1})
and (\ref{cinv2}) imply, for $z \neq x,y$, that
\begin{eqnarray}
\label{trace}
g^{\alpha \be}(z) G'{}_{\!\Delta}{}^i{}_{j,\alpha\beta}(x,y;z)
 &=& 0 \, , \\
\label{cons}
\nabla^\alpha_z G'{}_{\!\Delta}{}^i{}_{j,\alpha\beta}(x,y;z) &=& 0 \, .
\end{eqnarray}

Restricting to  flat space, 
\begin{equation}
\oG'{}_{\!\Delta}{}^i{}_{j,\alpha\beta}(x,y;z) =
G'{}_{\!\Delta}{}^i{}_{j,\alpha\beta}(x,y;z) \Big |_{g=\delta} \, ,
\label{Gvar2}
\end{equation}
which depends only on $x-z, y-z$,
is also strongly constrained by conformal invariance. It
may be expressed in the same form as was obtained for conformal
three point functions in \cite{OP,EO}, giving
\begin{equation} \label{var3p}
\oG'{}_{\!\Delta}{}^i{}_{j,\alpha\beta}(x,y;z) = -
\oG_\Delta{}^{i}{}_{i'}(x-z) \oG_\Delta{}^{j'}{}_{j}(z-y)
P^{i'}{}_{j'}{}_{, \,  \alpha \be} (Z) \, .
\end{equation} 
$P$ is a tensor symmetric and traceless in $(\alpha \be)$
which by conformal invariance depends only on
 the conformal vector $Z_\mu$. This transforms as a vector at $z$
and is defined by
\begin{equation} \label{Zdef}
Z_{ \mu } = \frac{(x-z)_{\mu}}{(x-z)^2} - \frac{(y-z)_{\mu}}
{(y-z)^2} \; \mbox{,}
\hspace{0.8cm} Z^2 = \frac{
(x-y)^2}{ (x-z)^2 (y-z)^2} \; \mbox{.} \label{Zdef2}
\end{equation} 
The tensor $P$ satisfies the homogeneity property
\begin{equation}
P^{i}{}_{j}{}_{, \, \alpha \be} (\lambda Z)
= \lambda^{s} P^{i}{}_{j}{}_{, \,  \alpha \be}
(Z) \, .
\end{equation}
Note that $ P^{i}{}_{j}{}_{, \,  \alpha \be}
(Z)$ has the crucial property that it does
not contain any factor $(Z^2)^{-n}$, $n=1,2, \dots$, 
since (\ref{varg}) implies that the only singular contributions for
two of the three points coincident involve $(x-z)$, $(y-z)$, but not
$(x-y)$.

By using results from \cite{EO} the conservation 
equation (\ref{cons}) can be simplified so 
as to constrain $P^{i}{}_{j}{}_{, \,  \alpha \be}$ alone. 
Instead of (\ref{var3p}) we may alternatively write using results from 
\cite{OP,EO}
\begin{equation}
\oG'{}_{\!\Delta}{}^i{}_{j,\alpha\beta}(x,y;z) =  - C_\Delta{}^{\! 2} \,
\frac{ I_{\alpha\epsilon}(z-x)I_{\beta\eta}(z-x) 
D^{j'}{}_{j}\Big(
I(x-y)\Big)}{\Big ( (z-x)^{2}\Big )^d \Big((x-y)^2 \Big)^{{1 \over 2 }(d-s)}
 } {\tilde P}^{i}{}_{j'}{}_{, \,  \epsilon \eta}
(X) \, .
\end{equation} 
where
\begin{equation}
{\tilde P}^{i}{}_{j}{}_{, \,  \alpha \beta}(X) = 
\frac{1}{(X^2)^{{1 \over 2 }(d+s)}} P^{i}{}_{j'}{}_{, \,  \alpha \be}
(X) D^{j'}{}_{j}\Big( I(X) \Big) \, ,
\label{tildeP}
\end{equation}
and
\begin{equation} \label{Xdef}
X_{ \mu } = \frac{(y-x)_{\mu}}{(y-x)^2} - \frac{(z-x)_{\mu}}
{(z-x)^2} \, .
\end{equation} 
Then on flat space (\ref{cons}) is equivalent to
\begin{equation} \label{consflat}
\pr_{\alpha} {\tilde P}^{i}{}_{j}{}_{, \,  \alpha \beta}(X) = 0 \, .
\end{equation}

\section{Second Order Conformal Operator on $k$-Forms}

\setcounter{equation}{0}

As an example for the operator $\Delta$ introduced in the previous section, we
now discuss the the conformal second order operator acting on $k$-forms
in $d$ dimensions which was constructed by Branson \cite{Branson}. 
We begin with some definitions. 

 If $\ome= (1/k!) \ome_{\mu_1 \cdots \mu_k} \d x^{\mu_1} \wedge
\cdots \wedge \d x^{\mu_k}$ is a $k$-form,
the exterior derivative $\d$ acting on $k$-forms 
and its adjoint $\de$, satisfying $\d^2=\de^2=0$, are defined by
\begin{eqnarray}
(\d \ome)_{\mu_1 \cdots \mu_{k+1}} &=& \sum\limits_{j=1}^{k+1}
(-1)^{j-1} \pr_{\mu_j} \ome_{\mu_1 \cdots \hat{\mu}_j \cdots
\mu_{k+1}} = (k+1) \pr_{[\mu_1} \ome_{\mu_2\cdots \mu_{k+1}]}\, , \label{d1}\\
(\delta \ome)_{\mu_1 \cdots \mu_{k-1}} &= & - \nabla^\lambda
\ome_{\lambda  \mu_1 \cdots \mu_{k-1}} \nonumber\\
&=& -  \frac{1}{\sqrt g}\,
g_{\mu_1 \nu_1} \cdots g_{\mu_{k-1} \nu_{k-1}} \pr_\lambda \Big( \sqrt g
g^{\lambda \tau} g^{\nu_1 \rho_1} \cdots g^{\nu_{k-1} \rho_{k-1}} 
\ome_{\tau \rho_1 \cdots \rho_{k-1}} \Big) \, ,
\label{del1}
\end{eqnarray}
where the hat $\hat{\mu}_j$ indicates that the corresponding index is
to be omitted. 
Following Branson we also define, with $R_{\mu \nu}$ and $R$ the 
Ricci tensor and scalar curvature,
\begin{eqnarray}
J & \equiv & \frac{1}{2(d-1)} R \, , \label{jb}\\
K_{\mu \nu} & \equiv & \frac{1}{(d-2)} \left( R_{\mu \nu} - J g_{\mu
\nu} \right) \, , \label{kb}
\end{eqnarray}
so clearly $K_{\mu \nu}$ is the same tensor as used for the definition of the
Weyl tensor in (\ref{Weylg}).
These transform under local Weyl rescalings as 
\begin{equation} \label{jkvar}
\delta_\si J = 2 \si J + \nabla^2 \sigma \, , \; \; \delta_\si K_{\mu\nu} =
\nabla_\mu \nabla_\nu \sigma \, .
\end{equation}

To construct a second order conformal differential operator acting on
$k$-forms we first consider the variation of $\de d$ and $d\de$ under
$\delta_\si g^{\mu\nu} = 2 \si
g^{\mu \nu}, \, \de_\si \sqrt g = - d\si \sqrt g$ using the explicit
dependence on the metric exhibited in (\ref{del1}) to give,
for $\gamma \equiv \half (d- 2 k )$,
\begin{eqnarray}
\delta_\si ( \delta \d \ome)_{\mu_1 \cdots \mu_k} &=& 
2 \gamma \si ( \delta \d \ome)_{\mu_1 \cdots \mu_k} \nonumber\\ & & 
{} +2(\gamma-1) (k+1) \nabla^\lambda \left ( \sigma \nabla_{[\lambda}
\ome_{\mu_1 \cdots \mu_k]} \right ) \nonumber\\
\delta_\si( \d \delta  \ome)_{\mu_1 \cdots \mu_k}
&=&  - 2 \gamma \big( \d \delta ( \si \ome) \big)_{\mu_1 \cdots
\mu_k} \nonumber\\ & & {} - 2  (\gamma+1) k  \nabla_{[\mu_1} \left ( \si
\nabla^\lambda  \ome_{|\lambda|\mu_2 \cdots \mu_k]} \right ) \; .
\end{eqnarray}
Using
\begin{eqnarray}
(k+1) \nabla^\lambda ( \si \nabla_{[\lambda} \ome_{\mu_1 \cdots
\mu_k]} ) =  \nabla^\lambda ( \si \nabla_{\lambda} \ome_{\mu_1 \cdots
\mu_k} )   - k  \nabla^\lambda ( \si \nabla_{[\mu_1} \ome_{|
\lambda | \mu_2 \cdots
\mu_k]} )
\end{eqnarray}
we find
\begin{eqnarray}
\lefteqn{ \delta_\si \Big( (\gamma+1) ( \delta \d \ome)_{\mu_1 \cdots
\mu_k} + (\gamma - 1) ( \d \delta  \ome)_{\mu_1 \cdots \mu_k} )
\Big) \hspace{3cm} }  \nonumber \\
&=& 2 \gamma (\gamma+1) \si \big( \delta \d \ome \big)_{\mu_1 \cdots
\mu_k} + 
2 \gamma (\gamma-1) \big ( \d \delta (\si \ome) \big)_{\mu_1 \cdots \mu_k} 
\nonumber\\ & & {} - 2 (\gamma-1)(\gamma+1) k \Big[ \nabla^\lambda ( \si
\nabla_{[\mu_1} \ome_{|\lambda| \mu_2 \cdots \mu_k]} ) + \nabla_{[\mu_1
} ( \si \nabla^\lambda \ome_{|\lambda| \mu_2 \cdots \mu_k]} ) \Big]
\nonumber \\ & & {} +2 ( \gamma -1) (\gamma+1) \nabla^\lambda ( \si
\nabla_\lambda  \ome_{\mu_1 \cdots \mu_k} )  \, . \label{dfvar}
\end{eqnarray}
Single derivatives acting on $\sigma$ are then eliminated by virtue of the
identity  
\begin{eqnarray}
\nabla^\lambda \si \nabla_\mu + \nabla_\mu \si \nabla^\lambda &=& 
 \half ( \nabla^\lambda \nabla_\mu + \nabla_\mu \nabla^\lambda ) \si
\nonumber\\ & &  {} - \half \si ( \nabla^\lambda \nabla_\mu + \nabla_\mu
\nabla^\lambda  ) - ( \nabla_\mu \nabla^\lambda \si )  \; ,
\end{eqnarray} 
so that we obtain for the the conformal variation of the operator
\begin{eqnarray}
\D^{(k)}  & \equiv &
(\gamma+1)  \delta \d + (\gamma - 1)  \d \delta  \nonumber \\
& = & (\gamma+1) ( \delta \d + \d \delta ) - 2 \d \delta \, ,
\label{Dk} 
\end{eqnarray}
from (\ref{dfvar}) the result
\begin{eqnarray}
\delta_\si ( \D^{(k)} \ome)_{\mu_1 \cdots \mu_k}
&=& (\gamma+1) \si ( \D^{(k)} \ome)_{\mu_1 \cdots \mu_k} -
(\gamma-1) ( \D^{(k)} \si  \ome)_{\mu_1 \cdots \mu_k}
\nonumber\\ & & {}+ 2 (\gamma-1) (\gamma+1) k ( \nabla_{[\mu_1}
\nabla^\lambda \si)  \ome_{|\lambda| \mu_2 \cdots \mu_k]}
\nonumber\\ & & {} -  (\gamma-1)(\gamma+1) \nabla^2 \si \, \ome_{\mu_1
\cdots \mu_k} \; .
\end{eqnarray}
The last two terms involving second derivatives acting on $\si$ 
can be cancelled  
by terms linear in $J$ and $K_{\mu}{}^\nu$, using the results for their
variation exhibited in
(\ref{jkvar}). The resulting conformally covariant differential operator
\begin{equation}
\Delta^{(k)} =  \D^{(k)}   +
 (\gamma + 1)(\gamma -1)  \,    \big( J \, -2 k \, K
\big)
\label{Delk} 
\end{equation}
where 
\begin{equation}
\big( K \ome \big){}_{\mu_1 \cdots \mu_k} \equiv \,
K_{[\mu_1}{}^{\, \nu} 
\ome_{| \nu | \mu_2  \cdots \mu_k] }  \, .
\end{equation}
is identical with the  formula for $\Delta^{(k)}$  constructed by Branson.
{}From the above calculations this has the conformal variation as in (\ref{sd})
for $s=2$,
\begin{equation} \label{varformop}
\delta_\si \Delta^{(k)} = (\gamma  +1) 
\si \Delta^{(k)} - (\gamma -1) \Delta^{(k)} \si  \, .
\end{equation}
On $0$-forms, or scalar fields, $\de \to 0$ and therefore
\begin{equation} \label{scalopf}  
\Delta^{(0)} = \half (d+2) \left ( \delta \d + \half (d-2) J \right )
= \half (d+2) \Delta_2
\end{equation}
the operator reduces to the well known conformal differential
operator defined in (\ref{2dop}).

An alternative expression of the existence of the conformal differential
operator acting on $k$-forms $\Delta^{(k)}$ given by (\ref{Delk}) 
may be found in terms of the $d$-dimensional
action for the $k$-form $\A_{\mu_1 \dots \mu_k}$ given by
\begin{eqnarray}
S^{(k)}(g,\A) &=& \, \half \, 
\int \! \d^d x\,  \sqrt g \, \Big [ (\gamma + 1) \,
(\d \A){\cdot (\d \A)} + (\gamma - 1) \,
(\delta \A){\cdot (\delta \A)}  \nonumber \\
&& \qquad\qquad \ {}+ (\gamma + 1)(\gamma - 1) ( J \,\A{\cdot \A} - 2k\, \A 
{\cdot (K\A)}) \Big ] \, ,
\label{Sk}
\end{eqnarray}
where for $X,Y$ $k$-forms we define,
\begin{equation}
X{\cdot Y} = \frac{1}{k!} \, X^{\mu_1 \dots \mu_k}Y_{\mu_1 \dots \mu_k} \, .
\end{equation}
The result (\ref{varformop}) is then equivalent to
\begin{equation}
\delta_\si S^{(k)}(g,\A) = 0 \quad \mbox{if} \quad \delta_\si \A = 
(\gamma - 1) \, \si \A \, .
\end{equation}
It is convenient to define
\begin{equation}
2 \frac{\delta}{\delta g^{\alpha\beta}} S^{(k)}(g,\A) = \sqrt g \,
T^{(k)}{}_{\!\!\! \alpha\beta} \, , \quad
\oT^{(k)}{}_{\!\!\! \alpha\beta} = T^{(k)}{}_{\!\!\! \alpha\beta} 
\Big |_{g=\delta} \, .
\label{Svar}
\end{equation}
The calculation of the flat space  expression for
$\oT^{(k)}{}_{\!\!\! \alpha\beta}$ is straightforward
given the explicit form of $\d$ in (\ref{d1}), which is metric independent, 
and $\delta$
in (\ref{del1}) but the detailed formula is lengthy and so it is relegated to
appendix A.4. From this we may find
\begin{eqnarray} \!\!\!\!\!\!\!\!\!\!
\oT^{(k)}{}_{\!\!\! \alpha\alpha} &=&  - (\gamma-1) \, \A{\cdot (\oD^{(k)} \A)}
\, , \nonumber \\
\!\!\!\!\!\! \pr_\alpha \oT^{(k)}{}_{\!\!\! \alpha\beta} 
&=& - {\textstyle\frac{1}{k!}} \, (\d \A)_{\beta \mu_1\dots \mu_k}\,
(\oD^{(k)} \A)_{\mu_1\dots \mu_k}   - {\textstyle\frac{1}{(k-1)!}}\,
\A_{\beta \mu_1\dots \mu_{k-1}} \, (\delta \oD^{(k)} \A)_{\mu_1\dots \mu_{k-1}}
\, ,
\label{tcon}
\end{eqnarray}
with $\oD^{(k)}$ the flat space restriction of $\D^{(k)}$, as in (\ref{Dk}),
and hence of the conformal differential operator $\Delta^{(k)}$.

According to the definition (\ref{dgreen}), the Green function of the
operator $\Delta^{(k)}$ is defined by
\begin{equation}
{\ts\sqrt{g(x)}}  \Big( \Delta_x{}^{\!\! (k)} G^{(k)}  
\Big){}_{\mu_1 \cdots \mu_k,}{}^{\! \nu_1 \cdots \nu_k} (x,y) = 
\E^A{}_{\!\!\mu_1 \cdots \mu_k,}{}^{\!\nu_1\cdots \nu_k}\, \delta^d(x-y) \, ,
\label{greenk}
\end{equation}
for $\E^A$ the projector onto totally antisymmetric $k$-index tensors,
\begin{equation}
\E^A{}_{\!\!\mu_1 \cdots \mu_k,}{}^{\!\nu_1\cdots \nu_k} =
\delta_{[\mu_1}{}^{\!\! \nu_1}\dots \delta_{\mu_k]}{}^{\! \nu_k} \, .
\end{equation}
In appendix A.2 we calculate this Green function 
for general $d$ and $k$ on flat space when $g_{\mu\nu}=\delta_{\mu\nu}$
and we may identify up and down indices. The result is
\begin{equation} \label{greenformop}
\oG^{(k)}{}_{\!\!\mu_1 \cdots \mu_k, \nu_1 \cdots \nu_k}(x)
= T_d \, {\cal I}^A{}_{\!\!\mu_1 \cdots \mu_k, \nu_1
\cdots \nu_k} (x) \frac{1}{x^{d-2}} \, ,
\end{equation} 
where
\begin{equation}
T_d = \frac{1}{S_d} \, \frac{1} {2(\gamma-1)(\gamma+1)} \, ,
\qquad S_d = \frac{2\pi^{d/2}} {\Gamma(\half d)} 
\end{equation}
and $ {\cal I}^A{}_{\!\mu_1 \cdots \mu_k, \nu_1 \cdots \nu_k} (x) $ is
the appropriate  $k$-form representation for inversions and is given by
\begin{equation}
{\cal I}^A{}_{\!\mu_1 \cdots \mu_k, \nu_1 \cdots \nu_k} (x) = 
{\cal E}^A{}_{\!\mu_1 \cdots \mu_k, \si_1 \cdots \si_k} I_{\si_1 \nu_1}
(x) \cdots I_{\si_k \nu_k} (x) \, .
\label{IA}
\end{equation}
Note that this Green function does not exist if $\gamma=\pm 1$.
If $\gamma = 1$, or $k= \half(d-2)$, then $\Delta^{(k)} =2\delta \d$ and
there are solutions satisfying $\Delta^{(k)} \psi = 0$ for $\psi= \d\phi$.
When $\gamma = -1$, or $k= \half(d+2)$, $\Delta^{(k)} = - 2\d \delta $ so
the operators are not invertible.  
The result for the Green function (\ref{greenformop}) is exactly
of the  form expected
from the general results for conformal two point functions discussed
in section 2 as given in (\ref{ggreen}). 

The conformal variation of the operator $\Delta^{(k)}$ 
defined in (\ref{Delk}) as given by (\ref{varformop}) remains
unchanged if we add a term involving the Weyl tensor
\begin{equation} \label{weyladd}
({\widetilde \Delta}^{(k)} \ome)_{\mu_1 \cdots \mu_k} = 
(\Delta^{(k)} \ome)_{\mu_1 \cdots \mu_k}
+t \,  C_{[\mu_1 \mu_2}{}^{  \si
\rho} \ome_{|\si \rho| \mu_3 \cdots \mu_k]} 
\end{equation}
for any real $t$ since $\delta_\si C_{\mu \nu}{}^{\si\rho} =
2 \si \, C_{\mu \nu}{}^{\si\rho}$.

We may now calculate the
variation of the Green function ${\widetilde G}^{(k)}$ for this operator
with respect to the metric using the formula (\ref{varg}). The
variation of the differential ope\-rator ${\widetilde\Delta}^{(k)}$ may be 
obtained by varying the expressions for 
$(\d \delta \ome)_{\mu_1 \cdots \mu_k}$ 
and $(\delta \d  \ome)_{\mu_1 \cdots \mu_k}$ using (\ref{d1},\ref{del1}),
which give the explicit dependence on the metric, 
and also calculating directly the variation of $J, K_{\mu\nu}$ defined in
(\ref{jb},\ref{kb}), but is equivalently given by using the 
expressions obtained from (\ref{Svar}).
Assuming the result (\ref{var3p}) simplifies the calculation
considerably, since it is sufficient to only determine the most singular 
terms as $z\to y$ which arise from terms with two derivatives
acting on $G^{(k)}{}_{\! \rho_1 \cdots \rho_k , \nu_1 \cdots \nu_k } (z,y)$,
 and we then find, with definitions analogous to (\ref{Gvar1},\ref{Gvar2}),
\begin{eqnarray} \label{dgf}
\lefteqn{\!\!\!\!\!\!\!\!\!
{\topcirc{\widetilde{G}}}{}^{(k)\prime}
{}_{\!\!\! \mu_1 \cdots \mu_k , \nu_1 \cdots \nu_k,\alpha\beta } (x,y;z) 
 = - \oG^{(k)}{}_{\!\! \mu_1 \cdots \mu_k ,
\si_1 \cdots \si_k}(x-z) 
\oG^{(k)}{}_{\!\!\rho_1 \cdots \rho_k, \nu_1 \cdots \nu_k}(z-y) \,
P_{\si_1 \cdots \si_k, \rho_1 \cdots \rho_k ,\alpha\beta}(Z)}
\nonumber\\
&& \!\!  {} -  2t \, 
\E^C{}_{\!\! \alpha\gamma\de\beta,\kappa\lambda\eps\eta}
\pr^z{}_{\! \gamma} \pr^z{}_{\! \de} \left (
\oG^{(k)}{}_{\!\!\mu_1 \cdots \mu_k ,\kappa\lambda \si_3 \cdots
\si_k}(x-z) 
\oG^{(k)}{}_{\!\!\eps\eta \si_3 \cdots \si_k ,\nu_1 \cdots \nu_k}(z-y)
\right ) \, , \hspace{1cm}
\end{eqnarray}
where $\E^C$ is a projection operator onto tensors with Weyl symmetry
defined in appendix A.1 and $\oG^{(k)}$ is as in (\ref{greenformop}). The
second line corresponds to the additional term in (\ref{weyladd}). In the
first line from direct calculation we find
\begin{eqnarray}
\lefteqn{P_{\mu_1 \cdots \mu_k ,
\nu_1 \cdots \nu_k,\alpha\beta} (Z) }\nonumber \\
& = &
d \frac{(\ga+1)(\ga-1)}{(d-1)(d-2)} \bigg[\, 2k (kd - d + k)
  \, \E^A{}_{\!\! \mu_1 \cdots \mu_k,\eps \rho_2 \cdots \rho_k} 
\E^A{}_{\!\! \eta \rho_2 \cdots \rho_k, \nu_1 \cdots \nu_k}
\E^T{}_{\!\! \eps \eta,\alpha\beta} Z^2 \nonumber\\
& & \hspace{3cm} {}+  4k(k-1)(d+1) \ga 
\E^A_{\mu_1 \cdots \mu_k, \kappa \eps \rho_3 \cdots \rho_k}
\E^A_{\lambda \eta \rho_3 \cdots \rho_k, \nu_1 \cdots \nu_k}
\E^T_{\ka \la, \alpha \beta} Z_\eps Z_\eta \nonumber\\ 
& & \hspace{3cm} {}+   k (d^2-2kd-d-2k+2)
\, \E^A{}_{\!\! \mu_1 \cdots \mu_k 
,\eps \rho_2 \cdots \rho_k} \E^A{}_{\!\! \eta \rho_2 \cdots \rho_k ,
\nu_1 \cdots \nu_k} \nonumber\\ & & \hspace{8.5cm} \times \,
\big ( \E^T{}_{\!\! \eps \kappa,\alpha\beta} Z_\eta +
\E^T{}_{\!\! \eta \kappa,\alpha\beta} Z_\eps \big ) Z_\kappa \nonumber\\
& &\hspace{3cm} {} -{\textstyle \frac{1}{2}}(d^2-2kd-4d+4) 
\, \E^A{}_{\!\!\mu_1 \cdots \mu_k  ,\nu_1 \cdots \nu_k } 
\big( Z_\alpha Z_\beta 
- {\textstyle{ 1 \over d} \de_{\alpha\beta}} Z^2\big ) \, \bigg] \, ,
\end{eqnarray}
with $\E^T{}_{\!\! \eps \eta,\alpha\beta} = \half ( \de_{\eps\alpha}
\de_{\eta\beta} + \de_{\eps\beta} \de_{\eta\alpha}) - \frac{1}{d}\de_{\eps\eta}
\de_{\alpha\beta}$ the projector onto symmetric traceless tensors. The 
tracelessness
equation (\ref{trace}) and conservation equation (\ref{cons}) follow
directly for flat space at non coincident points from the result (\ref{tcon}).
We have verified that this result is in accord with the conservation equation 
(\ref{consflat}) with $P \to {\tilde P}$ defined as in (\ref{tildeP}).

For the case $d=4$ and $k=2$ giving $\gamma = 0$ 
the Green function and its variation have been used in \cite{EO}
to construct an anomaly-free contribution to the effective action involving
the field strength tensor of a background gauge field. For the scalar
operator $\Delta_2$ defined in (\ref{2dop}) then, 
making use of (\ref{scalopf}), we have as a special
case of the above
\begin{eqnarray}
\oG_2(x) & = & 
\frac{1}{(d-2)S_d} \, \frac{1}{x^{d-2}} \, , \label{0form}\\
\oG{}'{}_{\!2,\alpha\beta}(x,y;z) &=& 
\frac{d(d-2)^2}{4(d-1)} \, \oG_2(x-z) \oG_2(y-z) \Big ( Z_\alpha Z_\beta
- { 1 \over d} \de_{\alpha\beta} Z^2\Big ) \, . \label{0form2}
\end{eqnarray}
In this case the conservation equation (\ref{consflat}) simply reduces to
\begin{equation}
\pr_\alpha \bigg ( \frac{1}{(X^2)^{\frac{1}{2}d + 2}}
\Big ( X_\alpha X_\beta - { 1 \over d} \de_{\alpha\beta} X^2\Big )\bigg) = 0 \, .
\label{Xcons}
\end{equation}

\section{Conformal Invariants}

\setcounter{equation}{0}

Before constructing a further conformal differential operator 
it is convenient to consider possible scalar fields constructed in terms
of the metric which transform homogeneously under  local Weyl
rescalings of the metric. It is trivial to construct such scalars in terms
of the Weyl tensor defined  in (\ref{Weylg}), since for $\de_\si g^{\mu\nu}
= 2\si g^{\mu\nu}$ this transforms as 
\begin{equation} \label{wscaling}
\delta_\si C_{\mu \si \rho \nu} =  -2 \si \,
C_{\mu \si \rho \nu} \, .
\end{equation}
Thus $F$ defined in (\ref{F}) transforms as $\de_\si F = 4\si F$. Using
the symmetry and trace conditions
\begin{equation}
 C_{\mu \alpha \beta \nu}  = C_{[\mu \alpha] [\beta \nu ]} 
\mbox{,} \; \; \;
 C_{\mu [\alpha \beta \nu ]} = 0 \mbox{,} \; \; \;
g^{\mu\nu} C_{\mu \alpha \beta \nu}  = 0 \;
\mbox{,} \label{Weyl}
\end{equation}
there are two such scalars cubic in the Weyl tensor,
\begin{eqnarray} \label{omegas}
\Omega_1 &=& C^{\mu \si}{}_{ \rho \nu} C^{\rho \nu}{}_{\alpha \beta}
C^{\alpha \beta}{}_{\mu \si} \, ,
\\
\Omega_2 &=& C^{\mu \si \rho \nu} C_{\nu \beta \si \alpha}
C^{\alpha}{}_\rho{}^\beta {}_\mu  \, ,
\end{eqnarray}
which, from (\ref{wscaling}), satisfy $\delta_\si \Omega_1 = 6 \si \Omega_1$,
$\delta_\si \Omega_2 = 6 \si \Omega_2$. For general $d$ these are independent
but when $d=4$, $\Omega_1 = 4 \Omega_2$ since
\begin{equation}
5\, C^{\mu \si}{}_{[\rho \nu} C^{\rho \nu}{}_{\alpha \beta}
C^{\alpha \beta}{}_{\mu ] \si} = \Omega_1 - 4 \Omega_2 \, .
\label{4d}
\end{equation}

As shown by Fefferman and Graham \cite{Fefferman} there is an additional
scalar $H$ satisfying
\begin{equation} \label{Hvar}
\de_\si H = 6 \si H \, .
\end{equation}
We here verify the existence of a scalar satisfying (\ref{Hvar})  
since in the physics literature results have been given only
for Weyl rescaling invariant integrals over all space in six dimensions
\footnote{See the second paper in \cite{WZ} or in \cite{Deser} with
corrections in \cite{Deser2}}.
For this purpose it is useful to define the Cotton tensor 
\begin{equation} 
\tilde{C}_{\beta \gamma \delta} = \nabla_\delta K_{\beta \gamma }
- \nabla_\gamma K_{\beta \delta } \, ,
\end{equation}
with $K_{\beta \gamma}$ as in (\ref{kb}).
It satisfies the identities
\begin{equation}
\tC_{\beta \gamma \de} = - \tC_{{\be {\de} \ga}} \, \;\; \mbox{and}
\hspace{1cm} \tC_{\beta \gamma \de} + \tC_{ \gamma \de \beta} + \tC_{
\de \be  \gamma } = 0 \, ,
\end{equation}
and by virtue of the standard Bianchi identities
\begin{equation} \label{Bianchi}
C_{\alpha \beta [\ga \de ; \eps ]} = \tC_{\alpha [\ga \de} g_{\eps] \be}
- \tC_{\beta [\ga \de} g_{\eps] \alpha} \, .
\end{equation}
{}From this result we may easily obtain
\begin{equation} \label{cctilde}
\nabla^\alpha C_{\alpha \beta \gamma \delta} = - (d-3) \tC_{\be \ga \de}
\, .
\end{equation}
For the variation of the Cotton tensor under local Weyl rescalings we have
\begin{equation}
\de_\si \tC_{\be \ga \de} = \pr_\alpha \si \, C^{\alpha}{}_{ \be \ga \de}
\, .
\end{equation}

The formula of Fefferman and Graham \cite{Fefferman} can be written as
\begin{equation} \label{HFG}
H = V^{\alpha\beta\gamma\delta\epsilon}V_{\alpha\beta\gamma\delta\epsilon}
+ 16 \, \tC^{\ga \de \eps} \tC_{\ga \de \eps} 
+ 16\, C^{\alpha \ga \de \eps} ( - \nabla_\alpha \tC_{\ga \de \eps}
+  K_{\alpha \be} C^{\be}{}_{\ga \de \eps} ) \, ,
\end{equation}
where 
\begin{equation}
V_{\alpha \be \ga \de \eps} \equiv \nabla_{\alpha}
C_{\be \ga \de \eps}  + 2 ( g_{\alpha [ \beta} \tC_{ \ga ] \de \eps}
+ g_{\alpha [ \de} \tC_{ \eps ] \be \ga}    )  \, .
\end{equation}
Using (\ref{cctilde}),
the invariant (\ref{HFG}) may alternatively be written as
\begin{eqnarray} \label{h}
H &=&  \nabla^\alpha C^{ \beta \gamma \delta \eps}
\nabla_\alpha C_{ \beta \gamma \delta \eps}
- 4(d-10) \tC^{\ga \de \eps} \tC_{\ga \de \eps} \nonumber\\
& & + \, 16 \, ( - C^{\alpha \ga \de \eps} \nabla_\alpha \tC_{\ga \de \eps}  
+ \, K_{\alpha \be} C^{\alpha \ga \de \eps} C^{\be}{}_{\ga \de \eps} ) \, .
\end{eqnarray}
The variation under local Weyl rescalings may be calculated with the aid of
\begin{eqnarray}
\lefteqn{\de_\si ( \nabla^\alpha C^{ \beta \gamma \delta \eps}
\nabla_\alpha C_{ \beta \gamma \delta \eps})  = 
6 \si \, \nabla^\alpha C^{ \beta \gamma \delta \eps}
\nabla_\alpha C_{ \beta \gamma \delta \eps}}  \nonumber \\
&& {} + \pr_\alpha \si \left ( 4 \nabla^\alpha C^{ \beta \gamma \delta \eps}
C_{ \beta \gamma \delta \eps} + 8 \nabla^\beta C^{ \alpha \gamma \delta \eps}
C_{ \beta \gamma \delta \eps} - 8 C^{ \alpha \gamma \delta \eps}
\nabla^\beta C_{ \beta \gamma \delta \eps}\right ) \, ,
\end{eqnarray}
and (\ref{cctilde}),
which then ensures (\ref{Hvar}).

Independently, Parker and Rosenberg \cite{Rosenberg} found an equivalent scalar 
which can be written in the form
\begin{equation}
\Omega = \frac{d-10}{d-2} \Big ( \nabla^\alpha C^{ \beta \gamma \delta \eps}
\nabla_\alpha C_{ \beta \gamma \delta \eps} - 4(d-2) 
\tC^{\ga \de \eps} \tC_{\ga \de \eps} \Big ) + \frac{4}{d-2}
\Big ( \nabla^2 - 4J \Big ) C^{ \beta \gamma \delta \eps}
C_{ \beta \gamma \delta \eps} \, .
\label{Omega}
\end{equation}
With standard results for the commutator
\begin{eqnarray}
C^{ \alpha\gamma \delta \eps}[\nabla_\alpha , \nabla_ \beta] 
C^\beta{}_{\gamma \delta \eps} &=& -(d-2) 
K_{\alpha \be} C^{\alpha \ga \de \eps} C^{\be}{}_{\ga \de \eps} - J \,
C^{\beta\ga \de \eps} C_{\be\ga \de \eps} + \Omega_1 + 2\Omega_2 
\nonumber \\
&=& -(d-2) C^{\alpha\gamma\delta \eps} \nabla_\alpha \tC_{\gamma \delta \eps}
- \half\, C^{\alpha\gamma\delta \eps} \nabla^2 C_{\alpha\gamma\delta \eps} \, ,
\label{CCd}
\end{eqnarray}
where in the second line we use $C^{\alpha\gamma \delta \eps}\nabla^\beta
\nabla_\alpha C_{\beta\gamma \delta \eps}  = \half C^{\alpha\gamma \delta\eps}
\nabla^2 C_{\alpha\gamma \delta\eps} - C^{\alpha\gamma \delta\eps} \nabla_\eps
{\tilde C}_{\de \gamma\alpha}$, from (\ref{Bianchi}), and (\ref{cctilde})
for $\nabla^\beta C_{\beta\gamma \delta\eps}$.
 Using (\ref{CCd}),
with $H$ as in (\ref{h}), we then find
\begin{equation}
H = \Omega + \Omega_1 + 2 \Omega_2 \, .
\end{equation}
It is perhaps worth noting that the formula (\ref{Omega}) can also be written
in the form
\begin{eqnarray}
\Omega &=& - \frac{1}{d-2} \Big ( (d-10)C^{ \beta \gamma \delta \eps} \nabla^2
C_{ \beta \gamma \delta \eps} + 16J \, C^{ \beta \gamma \delta \eps} 
C_{ \beta \gamma \delta \eps} \Big ) \nonumber \\
&&{}- 4(d-10) \, \tC^{\ga \de \eps} \tC_{\ga \de \eps} + \half \nabla^2
( C^{ \beta \gamma \delta \eps} C_{ \beta \gamma \delta \eps} ) \, .
\end{eqnarray}

\section{Conformal Operator on Weyl Tensor Fields}

\setcounter{equation}{0}

As a further example of a conformally covariant second order differential
operator we now construct a second order differential operator
acting on tensor fields, $\C_{\mu \si \rho \nu}$, with Weyl symmetry, i.e.
they satisfy the symmetry and traceless conditions of the Weyl tensor
exhibited in (\ref{Weyl}). We assume that $\C_{\mu \si \rho \nu}$ is to
be regarded as metric independent but also to transform with  local rescalings
of the metric $\de_\si g^{\mu \nu} = 2\si g^{\mu \nu}$ according to
\begin{equation} \delta_\si \C_{\mu \si \rho \nu} \, = \, - \, k \, \si \,
\C_{\mu \si \rho \nu}  \, , \label{kdc1}
\label{Csi}
\end{equation}
with $k$ an arbitrary number.

In order to find such a conformally covariant covariant differential operator, 
we construct a reparametrisation action of second order in $\C$ and then
determine the restrictions necessary to ensure local scale invariance
(a similar procedure was followed in a recent 
paper  by  O'Raifear\-taigh, Sachs and Wiesendanger \cite{Raifertaigh}). 
We therefore first 
consider the $d$-dimensional action with two derivatives, which by using
the symmetry properties of $\C_{\mu \si \rho \nu}$, can be reduced to
\begin{equation} \label{w2}
S_0[g,\C] \, = \, {\half} \int \! \d^d x \, 
\sqrt g \, \Big[ a \, \nabla^\alpha \C^{\mu \si \rho \nu}
\nabla_\alpha \C_{\mu \si \rho \nu} \, + \, b \, \nabla_\mu 
\C^{\mu \si \rho \nu} \nabla^\alpha \C_{\alpha \si \rho \nu} \Big] \, .
\end{equation}
The two terms in (\ref{w2}) with coefficients $a$ and $b$ are the only
possible independent 
scalars involving $\C_{\mu \si \rho \nu}$.
In the following we determine the values for the two
coefficients $a$ and $b$ necessary for a conformally covariant second
order differential operator acting on $\C_{\mu \si \rho \nu}$ by
requiring $\delta_\si S^C[g,\C] = 0$, where  $S^C[g,\C]$  
contains further curvature dependent terms in addition to $S_0[g,\C]$.
It is easy to see that for the action $S_0$ to be invariant for constant $\si$
then we must have 
\begin{equation} \label{kdc2}
k = - \half (d-10) \, .
\end{equation}

To calculate the variation for general $\si$ we make use of
\begin{eqnarray}
\de_\si \left( \nabla^\alpha \C^{\mu \si \rho \nu}
\nabla_\alpha \C_{\mu \si \rho \nu}  \right) & = & (10-2 k) \, \si \,
\nabla^\alpha \C^{\mu \si \rho \nu} \nabla_\alpha \C_{\mu \si \rho \nu} 
+ 2 (4-k) \pr_\alpha \si \,\nabla^\alpha \C^{\mu \si \rho \nu}
\C_{\mu \si \rho \nu}  \nonumber\\
& & {} + 8 \pr_\mu \si \, \nabla^\alpha \C^{\mu \si \rho \nu}
 \C_{\alpha \si \rho \nu}  
\, - \, 8 \pr_\alpha \si\, \nabla_\mu \C^{\mu \si \rho \nu}
\C^{\alpha}{}_{ \si \rho \nu} \, , \\
\de_\si  \nabla^\alpha \C_{\alpha \si \rho \nu}  & = &
(2-k)  \si \, 
 \nabla^\alpha \C_{\alpha \si \rho \nu}
\, + \, (5-k-d) \pr_\alpha \si \, \C^{\alpha}{}_{ \si \rho \nu} \, .
\end{eqnarray}
Using these results for (\ref{w2}) with (\ref{Csi},\ref{kdc2}) we obtain
\begin{eqnarray}
\de_\si S_0[g,\C] \, &=& \, {\half} \, \int \! \d^d x \sqrt g \, \Big[ \, 
a \Big( \half (d-2) \pr_\alpha \si  \nabla^\alpha
( \C^{\mu \si \rho \nu} \C_{\mu \si \rho \nu} )
- {} 16 \pr_\alpha \si \nabla_\mu 
\C^{\mu \si \rho \nu} \, \C^{\alpha}{}_{\si \rho \nu}
\nonumber\\ & & \hspace{2.3cm} - 8\, \nabla_\alpha \pr_\mu \si  \,
\C^{\mu \si \rho \nu}  \C^{\alpha}{}_{\si \rho \nu} \Big) {}
 -  b d\, \pr_\alpha \si  \nabla_\mu 
\C^{\mu \si \rho \nu}  \C^{\alpha}{}_{\si \rho \nu}
 \, \Big] \, . \nonumber\\ & & 
\end{eqnarray}
Hence if we choose
\begin{equation}
16 \, a = - {d} \,  b \, ,
\end{equation}
the terms involving single derivatives of $\si$ in $\de_\si S_0$ cancel
and now
\begin{equation}
\de_\si S_0[g,\C] 
=  - {\half}  a  \int \! \d^d x \sqrt g \, \Big[ \half
(d-2) \nabla^2 \si \, \C^{\mu \si \rho \nu}
\C_{\mu \si \rho \nu} \,
+ 8 \nabla_\alpha \pr_\mu \si \, \C^{\mu \si \rho \nu}
\C^{\alpha}{}_{ \si \rho \nu}   \, \Big] \, . \nonumber\\  
\end{equation}
These terms may be cancelled by the conformal variation of the
additional action 
\begin{equation}
S_1[g,\C] =  \half a  \int \! \d^d x  \sqrt g \, \Big[ 
\half (d-2)J \, \C^{\mu \si \rho \nu} \C_{\mu \si \rho \nu}   +  8  
K_{\mu\alpha}\,
\C^{\mu \si \rho \nu} \C^{\alpha}{}_{ \si \rho \nu}  \Big] 
\label{w3}
\end{equation}
involving the curvature dependent terms $J$ and $K_{\mu \nu}$ defined in
(\ref{jb}) and (\ref{kb}). Hence, assuming now $a =  d/16, b=-1$,
we obtain a conformally invariant action
\begin{eqnarray} \label{sdeltac}
S^C[g,\C] & = & S_0[g,\C] + S_1[g,\C] 
\nonumber\\  & = & {\half} \, \int \! \d^d x  \sqrt g \,
\C^{\mu \si \rho \nu} ( \Delta^C \C )_{\mu \si \rho \nu} \,  \, ,
\end{eqnarray}
which defines the conformally covariant differential operator $\Delta^C$ 
where explicitly
\begin{eqnarray}  \label{deltac}
(\Delta^C \C )_{\mu \si \rho \nu} & = &  - {\textstyle{\frac{1}{16}}} d \, 
  \Big( \nabla^2
- \half (d-2) J \Big) \C_{\mu \si \rho \nu} \nonumber\\
& & {} + \E^C{}_{\!\!\mu \si \rho \nu,}{}^{\, \mu' \si' \rho' \nu'}   
\Big( \nabla_{\mu'} \nabla^\alpha + {\textstyle{\frac{1}{2}}} d \,  
 K_{\mu'}{}^{\alpha} \Big) 
\C_{\alpha \si' \rho' \nu'} \, ,
\end{eqnarray}
with $\E^C$ as  defined in appendix A.1.
As a consequence of $\de_\si S^C[g,\C]=0$, as well as 
(\ref{kdc1}) and (\ref{kdc2}), it must
satisfy
\begin{equation}
\de_\si \Delta^C =   \half (d-6) \, \si \Delta^C - \half (d-10)\,  \Delta^C
\si \, .
\end{equation}
For generality we could also add
$S_2[g,\C]$ containing the invariants
\begin{eqnarray} \label{s2gc}
S_2[g,\C] = \half \, \int \! \d^d x \sqrt g \, \Big [
c\, \C^{\mu \si \rho \nu} \C_{\nu \beta \si \alpha}
C^{\alpha}{}_\rho{}^\beta {}_\mu + e \, \C^{\mu \si \rho \nu} \C_{\rho \nu
\alpha \beta} C_{\mu \si}{}^{\alpha \beta} \Big ]\, ,
\end{eqnarray}
where $C_{\mu \ka \la \nu}$ is the standard Weyl tensor
defined in (\ref{Weylg}). Letting ${\tilde S}^C[g,\C] = S^C[g,\C] + S_2[g,\C]$
we therefore obtain a two parameter conformal differential operator
${\tilde \Delta}^C$ depending on $c$ and $e$,
\begin{eqnarray} \label{ce}
(\tilde{\Delta}^C \C )_{\mu \si \rho \nu} &=&  (\Delta^C \C)_{\mu
\si \rho \nu} \nonumber\\ & & {} + \E^C{}_{\!\!\mu \si \rho \nu,}{}^{ \mu'
 \si' \rho' \nu'} \Big( c \, C^\alpha{}_{\rho'} {}^\beta {}_{\mu'}
\C_{\nu' \beta \si' \alpha} 
 + \, e \, C_{\mu' \si' }{}^{\alpha \beta}
\C_{\rho' \nu' \alpha \beta}  \Big) \, .
\end{eqnarray}

The corresponding Green function is defined by
\begin{equation} \label{cgdef}   {\ts\sqrt{g(x)}} 
( \Delta_x{}^{\!\! C} G^C )_{\mu \si \rho \nu,}{}^{\alpha \ga\delta
\beta} (x,y) = \E^C{}_{\!\!\mu \si \rho \nu,}{}^{\alpha \ga\delta
\beta} \, \delta^d(x-y) \, .
\end{equation}
On flat space  this may be calculated 
in a similar fashion as previously by inverting the Fourier
transform of $\Delta^C$. From the results in appendix A.3
\begin{equation}
\oG^C{}_{\!\!\mu \si \rho \nu, \alpha \gamma \delta \beta}  (x) 
= \frac{1}{S_d}\,\frac{16}{ (d-4) (d-6)} \,
\frac{1}{ x^{d-2}} \, \I^C{}_{\!\!\mu \si \rho \nu, \alpha \gamma \de \beta}
(x) \, , \label{cgreen}
\end{equation}
where the relevant form for the inversion tensor is
\begin{equation}
\I^C{}_{\!\!\mu \si \rho \nu, \alpha \gamma \de \beta}(x)
= I_{\mu\epsilon}(x)I_{\si\kappa}(x)I_{\rho\lambda}(x)I_{\nu\eta}(x)
\E^C{}_{\!\!\epsilon \kappa \lambda \eta, \alpha \gamma \de \beta} \, .
\end{equation}
This is clearly in accord with the general expression expected by
conformal invariance given by (\ref{ggreen}) which thus provides
a consistency check on the derivation of the conformal operator $\Delta^C$.
Obviously from (\ref{cgreen}) $\Delta^C$ is  not invertible $d=4$ or $d=6$.

The lack of an inverse for $d=4,6$ is true more generally. To demonstrate
this we first note that $\Delta^C$ essentially vanishes identically if $d=4$.
For $d=4$ from the vanishing of totally antisymmetric five index tensors
we have
\begin{eqnarray} \label{4ac1}
0 &=& 5 \, \nabla^\alpha\C^{\mu \si}{}_{ [ \rho \nu} \nabla_\alpha 
\C_{\mu\si]}{}^{\rho \nu} \nonumber \\
&= &
\nabla^\alpha \C^{\mu \si \rho \nu} \nabla_\alpha \C_{\mu \si \rho \nu}
-2 \, \nabla^\alpha\C_{\alpha}{}^{\si\rho\nu} \nabla_\mu \C^{\mu}{}_{\si\rho\nu}
-2 \, \nabla^\alpha \C^{\mu \si \rho \nu} \nabla_\mu \C_{\alpha \si \rho \nu}
\nonumber \\
& \to & \nabla^\alpha \C^{\mu \si \rho \nu} \nabla_\alpha \C_{\mu \si \rho \nu}
- 4 \nabla_\alpha \C^{\alpha\si\rho\nu} \nabla^\mu \C_{\mu\si\rho\nu}
+2 \, \C^{\mu \si \rho \nu} [\nabla^\alpha, \nabla_\mu] \C_{\alpha
\si \rho \nu} \, , 
\end{eqnarray}
after discarding a total derivative and where for general $d$
\begin{eqnarray} \label{commu}
\C^{\mu \si \rho \nu} [\nabla^\alpha , \nabla_\mu ] \C_{\alpha
\si \rho \nu} &=&  (d-2) \, K_{\mu\alpha}\C^{\mu \si \rho \nu} 
\C^{\alpha}{}_{ \si \rho \nu}  - J
\, \C^{\mu \si \rho \nu}  \C_{\mu \si \rho \nu} \nonumber\\ & & 
{}- 
C^{\alpha\beta}{}_{\mu\si} \C^{\mu\si}{}_{\rho\nu}\C^{\rho\nu}{}_{\alpha\beta}
- 2\, C^{\alpha}{}_\rho{}^\beta {}_\mu
\C^{\mu \si \rho \nu} \C_{\nu \beta \si \alpha} \, .
\end{eqnarray}
Similarly we have, using $K_{\alpha}{}^\alpha = J$,
\begin{equation}
0 = 5 \, \C^{\mu \si}{}_{ [\rho \nu} K_{\alpha}{}^\alpha 
\C_{\mu \si]}{}^{ \rho \nu}
= -4K_{\mu\alpha} \, \C^{\mu \si \rho \nu}  
\C^{\alpha}{}_{ \si \rho \nu}
+ J\, \C^{\mu \si \rho \nu} \C_{\mu \si \rho \nu} \, .
\label{4ac2}
\end{equation}
In order to be able to apply these identities we write the action
$S^C$ given in (\ref{sdeltac}) in four dimensions as
\begin{eqnarray}
S^C[g,\C]\Big |_{d=4} &= & \half \int \! \d^4x  \sqrt g     \Big[
{\textstyle{\frac{1}{4}}}  \,  \nabla^\alpha \C^{\mu \si \rho \nu}  
\nabla_\alpha  \C_{\mu \si \rho \nu} -  \nabla_\mu \C^{\mu \si
\rho \nu}   \nabla^\alpha  \C_{\alpha \si \rho \nu}  
\nonumber\\
& & \hspace{2cm}{} + 2\, K_{\mu\alpha} \C^{\mu \si \rho \nu} 
\C^{\alpha}{}_{ \si \rho \nu}  + {\textstyle{\frac{1}{2}}} \, J  
\, \C^{\mu \si \rho \nu}  \C_{\mu \si \rho \nu}\Big ] \nonumber\\ 
& = & \half \int \! \d^4x  \sqrt g 
\Big [ {\textstyle{1\over 4}}\,
C^{\alpha\beta}{}_{\mu\si} \C^{\mu\si}{}_{\rho\nu}\C^{\rho\nu}{}_{\alpha\beta}
+ \half \, C^{\alpha}{}_\rho{}^\beta {}_\mu
\C^{\mu \si \rho \nu} \C_{\nu \beta \si \alpha}  
\Big] \, .
\end{eqnarray}
by virtue of (\ref{4ac1}), (\ref{commu}) and (\ref{4ac2}).
For $d=4$ the two ${\rm O}(C\C^2)$ terms are not independent, just as in
(\ref{4d}), so in order to obtain ${\tilde S}^C[g,\C]=0 $ and hence
$(\widetilde{\Delta}^C \C)_{\mu \si \rho \nu} = 0$ it is sufficient to take
\begin{equation}
c \, + 4 \, e = -{\textstyle{3\over 2}} \; .
\end{equation}

In six dimensions the lack of an inverse follows since the Weyl tensor
itself is a zero mode, i.e.
\begin{equation}
({\tilde \Delta}^C C)_{\mu\si\rho\nu} = 0 \quad  \ \mbox{if} \quad  \
c = - {\textstyle {3 \over 2}} \, , \quad  e = - {\textstyle {3 \over 4}} \, .
\label{zero}
\end{equation}
Applying $\Delta^C$ to the Weyl tensor is consistent for $d=6$ since
the transformation of $C_{\mu\si\rho\nu}$ in (\ref{wscaling}) matches that for
$\C_{\mu\si\rho\nu}$ in (\ref{kdc1})  as then $k=2$ from (\ref{kdc2}),
To verify (\ref{zero}) for this case we now use (\ref{CCd}) with (\ref{cctilde})
for $d=6$ to give
\begin{equation}
S^C[g,C]\Big |_{d=6}
= -{\textstyle \frac{3}{8}} 
\int \! \d^6 x  \sqrt g \, ( \Omega_1 + 2 \Omega_2)\,.
\end{equation}
This may be cancelled by $S_2[g,C]$, defined in (\ref{s2gc}), for the
above choices of the parameters $c,e$, and as ${\tilde S}[g,C]=0$ (\ref{zero})
must hold.

As in (\ref{Svar}) we may also define
\begin{equation}
2 \frac{\delta}{\delta g^{\alpha\beta}} S^C(g,\C) = \sqrt g \,
T^{C}{}_{\!\! \alpha\beta} \, , \quad
\oT^{C}{}_{\!\! \alpha\beta} = T^{C}{}_{\!\! \alpha\beta}
\Big |_{g=\delta} \, .
\label{SvarC}
\end{equation}
The calculation of the flat space  expression for
$\oT^{C}{}_{\!\! \alpha\beta}$ is again straightforward although tedious
from (\ref{deltac}) and (\ref{w2}), with the assumed values of $a,b$,
and also (\ref{w3}).
The explicit result is given in appendix A.4. From this we may obtain
\begin{eqnarray}
\oT^{C}{}_{\!\! \alpha\alpha} &=&  - \half(d-10) \, \C_{\mu\si\rho\nu}
\, (\oDel^{C}\C)_{\mu\si\rho\nu} \, , \nonumber \\
\pr_\alpha \oT^{C}{}_{\!\!\! \alpha\beta}
&=& - (\oDel^{C}\C)_{\mu\si\rho\nu} \, \pr_\beta  \C_{\mu\si\rho\nu}
+ 4 \, \pr_\alpha \big (\C_{\beta\si\rho\nu}\,
(\oDel^{C}\C)_{\alpha\si\rho\nu} \big ) \, ,
\label{tC}
\end{eqnarray}
with $\oDel^{C}$ the flat space restriction of $\Delta^{C}$.

Using (\ref{varg}) we may find the variation of the Green function defined in 
(\ref{cgdef}), in analogy to the $k$-form case (\ref{dgf}). As before, 
assuming the result to be of the form (\ref{var3p}), which in this case
gives
\begin{eqnarray}
\lefteqn{
{\topcirc{{G}}}{}^{C\prime}{}_{\!\! \mu \kappa \lambda \nu, \si \eps 
\eta \rho, \alpha \be} (x,y ; z) }\nonumber\\ 
&=& \! - \oG^C{}_{\!\! \mu \kappa \lambda \nu, \mu' \kappa' \lambda' \nu' }\,
(x-z) \oG^C{}_{\!\! \si \eps \eta \rho, \si' \eps' \eta' \rho'}(y-z) \,
P_{\mu' \kappa' \lambda' \nu', \si' \eps' \eta' \rho',\alpha \be}(Z) \, ,
\label{GCres}
\end{eqnarray}  
with $\oG^C$ given by (\ref{cgreen}), simplifies the calculation 
considerably. It is sufficient to determine the most singular terms 
as $z \rightarrow y$. Using (\ref{SvarC}) with the flat space result
(\ref{A20}) of appendix A.4 gives a relatively lengthy
expression for $P_{\mu' \kappa' \lambda' \nu', 
\si' \eps' \eta' \rho',\alpha \be}(Z)$ which may be found in appendix A.5.
An important consistency check is that the conservation equations 
(\ref{consflat}) are satisfied.

\section{Fourth Order Conformal Operators}

\setcounter{equation}{0}

So far we have only dealt with second order operators. In this section we
discuss some results for fourth order operators.

A fourth order
$d$-dimensional conformal differential operator acting on scalars  was
found  by Paneitz \cite{Paneitz}
 and almost simultaneously in $d=4$
by Riegert \cite{Riegert} and later 
independently by Eastwood and Singer \cite{Eastwood}. This has the form
\begin{equation}
\Delta_{4} =  \nabla^2 \nabla^2 + \nabla_\mu ( 4 K^{\mu\nu} - (d-2)
g^{\mu\nu} J ) \pr_\nu  + 
{\textstyle{\frac{1}{2}}}{(d-4)} \, M \, ,
\label{delta4}
\end{equation}
where $J$ and $K$ are given by (\ref{jb}) and (\ref{kb}) and 
\begin{equation}
M  = - \nabla^2 J
+ \textstyle{\frac{1}{2}} dJ^2 - 2 K^{\mu \nu} K_{\mu \nu} \ .  
\label{scalopnot}
\end{equation}  
This operator is  conformally covariant in the sense that
\begin{equation}
\delta_\sigma   \Delta_{4} =   \half (d+4)\, 
\sigma \Delta_{4} - \half (d-4)\, \Delta_{4} \sigma 
 \, . \label{scalop4}
\end{equation}

Subsequently $\Delta_4$ was generalised to  differential forms
by Branson and also by W\"unsch \cite{Branson,Wuensch}. This operator has
the form, with the same notation as in section 3,
\begin{equation}
\Delta^{(k)}_4 = \D^{(k)}_4
+ \mbox{curvature dependent terms} \, , \quad
\D^{(k)}_4 = (\gamma + 2 )\, \de \d \de \d + (\gamma -2 )\, \d \de \d \de \, ,
\label{D4}
\end{equation}
which satisfies
\begin{equation}
\de_\si \Delta^{(k)}_4  = (\gamma + 2)\, \si \Delta^{(k)}_4 - (\gamma -2 )\,
\Delta^{(k)}_4 \si \, .
\end{equation}
Acting on functions it reduces to the operator in (\ref{delta4})
\begin{equation} \label{scaldel}
\Delta^{(0)}_4 = \half (d+4) \Delta_4 \, .
\end{equation}
The Green function $G_4^{(k)}(x,y)$ in for $\Delta^{(k)}_4$
is defined similarly to
(\ref{greenk}). On flat space, this Green function may be calculated
by the same technique as used for the second order operators in appendices A.2
and A.3. First we note that in $\D^{(k)}_4$ in (\ref{D4}) may be 
rewritten as
\begin{equation}
\D^{(k)}_4 = ((\gamma + 2 ) ( \de \d + \d \de) - 4 \, \d \de )
( \de \d + \d \de) \, ,
\end{equation}
so that when acting on $k$-forms, the flat space
restriction of the operator $\Delta^{(k)}_4$ is easily seen to be
\begin{equation}
\Big( \oDel^{(k)}_4 \ome \Big){}_{\mu_1 \cdots \mu_k} \Big|_{g= \delta}
= (\gamma+2)\,  \pr^2 \pr^2 \ome_{\mu_1 \cdots \mu_k}
- 4k \, \pr^2 \pr_\lambda \pr_{[\mu_1} \ome_{|\lambda | \mu_2 \cdots \mu_k]} 
\end{equation}
By inverting the Fourier transform of this expression and transforming
back to position space with the help of 
(\ref{fouriert},\ref{fouriert2}) and (\ref{fouriert3}), 
we obtain for the Green function on
flat space 
\begin{equation}
\oG^{(k)}_4{}_{\mu_1 \cdots \mu_k,\nu_1 \cdots
  \nu_k } (x) = \frac{\Gamma(\half d -1)}{16
  \pi^{d/2}(\gamma -2)(\gamma+2)} \, \I^A{}_{\!\!\mu_1 \cdots \mu_k,\nu_1 \cdots
  \nu_k }(x)\, \frac{1}{x^{d-4}} \, . \label{G4k}
\end{equation}
Just as in (\ref{greenformop}) this involves the inversion tensor for
$k$-forms as required by conformal invariance.

For further discussion for simplicity
we restrict our attention to the Green
function $G_4(x,y)$ for the scalar operator defined in (\ref{delta4}). From
(\ref{scalop4}) this behaves under rescalings of the metric according to
\begin{equation}
\de_\si G_4(x,y) = \half (d-4) ( \si(x) + \si(y) ) G_4(x,y) \, .
\label{cvarG4}
\end{equation}
On restriction to flat space it is easy to see either directly or from
(\ref{G4k}) for $k=0$ using (\ref{scaldel}),  
\begin{eqnarray}
\oG_4(x) =  
\frac{\Gamma(\half d - 2)}{16 \pi^{d/2}} \, \frac{1}{x^{d-4}} =
\frac{1}{2(d-2)(d-4)}\, \frac{1}{S_d}\, \frac{1}{x^{d-4}} \, .
\label{Hdef}
\end{eqnarray}
The calculation of the variation of $G_4$ with respect to the metric
is straightforward by taking account of the metric dependence of $\Delta_4$,
noting that  $\nabla^2  = ({\sqrt g})^{-1} \pr_\mu \sqrt
g g^{\mu \nu} \pr_\nu $, and gives
\begin{eqnarray}
\oG{}'{}_{\!4,\alpha\beta}(x,y;z) &=& -\frac{1}
{d-1} \,  \bigg[ \,
- \quar (d-4) \, ( X \, \pr_\alpha
\pr_\beta \pr^2 Y +  \pr_\alpha \pr_\beta \pr^2 X \, Y)
\nonumber\\ & & \qquad\qquad {} 
+ \pr_\alpha \pr_\beta \pr_\rho X \, \pr_\rho
Y + \pr_\rho X \, \pr_\alpha \pr_\beta \pr_\rho  Y
\nonumber\\ & & \qquad\qquad
{} - \half (d+2) \,
 ( \pr_{(\alpha} X \pr_{\beta)} \, \pr^2 Y + \pr^2 \pr_{(\alpha} X \, \pr_{\beta)} Y )
\nonumber\\ & & \qquad\qquad
{} +  \half \, \de_{\alpha \beta} ( \pr^2 \pr_\rho X \, \pr_\rho Y + \pr_\rho X
\, \pr^2 \pr_\rho Y )  
\,   \nonumber\\ & & \qquad\qquad
{} + \frac{d(d+2)}{4(d-2)} \,\Big ( 
\pr^2 X \, \pr_\alpha \pr_\beta Y + \pr_\alpha \pr_\beta
X \, \pr^2 Y  - \frac{2}{d}\,
\de_{\alpha \beta} \pr^2 X \, \pr^2 Y \Big ) 
\nonumber\\ & & \qquad\qquad
{} - \frac{2d}{d-2} \, \Big ( 
\pr_\rho \pr_{(\alpha} X \, \pr_{\beta)}  \pr_\rho Y  -
\frac{1}{d}\, 
\de_{\alpha \beta } \pr_\si \pr_\rho X \, \pr_\si \pr_\rho Y \Big )
\, \bigg] \, , \label{G4var}
\end{eqnarray}
where we employ the abbreviations
\begin{eqnarray} \label{Habb}
X(z)  \equiv \oG_4(z-x) \, , \;\;\;\; 
Y(z)  \equiv \oG_4(z-y) \, .
\end{eqnarray}
The explicit calculation of the derivatives yields an expression
in agreement with the general result (\ref{var3p}),
\begin{equation}
\oG{}'{}_{\!4,\alpha\beta}(x,y;z) = - \frac{
d(d-2)(d-4)^2} {2 (d-1)}  \, \oG_4(x-z) \oG_4(y-z)  
\, \Big(Z_\alpha Z_\beta - \frac{1}{d} \de_{\alpha \beta} Z^2 \Big ) Z^2 \, .
\label{G4var2} 
\end{equation}
This satisfies the conservation condition (\ref{consflat}) since it reduces
just to (\ref{Xcons}) again.

\section{Discussion}

\setcounter{equation}{0}

A crucial motivation for this work was to use the Green functions of
conformally covariant differential operators to construct expressions
for the effective action depending on the metric which reproduce exactly
the required results for scale anomalies.
In four dimensions the operator $\Delta_{4}$ 
reduces to the operator introduced by Riegert
\begin{equation}
\Delta^{\rm {R}}  \equiv \Delta_{4}\Big | _{d=4} = 
\nabla^2 \nabla^2 + 2 \nabla_\mu 
( R^{\mu\nu} - {\textstyle{1\over 3}}g^{\mu\nu}R )\partial_\nu \, ,
\end{equation}
which gives  $\delta_{\sigma} (\sqrt g \Delta^{\rm R}) = 0$. 
Its  Green function $G^{\rm R}(x,y)=G_4(x,y)|_{d=4}$, which is therefore 
invariant under local
rescalings of the metric 
$\de_\si G^{\rm R}(x,y)=0$, was used by Riegert to construct possible
forms for the effective action. To demonstrate this we may note that
\begin{equation}
{\cal G} \equiv \sqrt g ( G - {\textstyle {2\over 3}} \nabla^2 R) \, ,
\end{equation}
with $G$ the Gau\ss -Bonnet term defined in (\ref{GB}),
has the conformal variation
\begin{equation}
\delta_{\sigma} {\cal G} = - 4 \sqrt g \Delta^{\rm R} \sigma \, ,
\end{equation}
and hence
\begin{equation} 
 \delta_{\sigma} \Sigma = \sigma \, , \qquad
\Sigma(x) = -{\textstyle \frac{1}{4}}
 \,\int \, {\rm d}^4y \, G^{\rm R}(x,y){\cal G} (y) \, .
\end{equation}
If we generalise (\ref{FG}) slightly to 
\begin{equation} 
g^{\mu \nu} \l T_{\mu \nu}\r =
 - \F - \beta_b \, G \quad \mbox{for} \quad \de_\si \F = 4 \F \, ,
\end{equation}
then a four dimensional non-local  action, analogous to the two dimensional 
Polyakov action
(\ref{polyakov}), which generates the trace anomaly under local rescalings of
the metric was constructed by Riegert \cite{Riegert} of the form,
\begin{eqnarray}
W_{{\rm Riegert}}[g]
&=  &   \int \!  {\rm d}^4x\sqrt g  \, \F(x) \Sigma(x) \nonumber\\
&&{} - {\textstyle\frac{1}{8}}\beta_b \int \!  {\rm d}^4x {\rm d}^4y  \,
{\cal G} (x)  \, G^{\rm R}(x,y) \, {\cal G}(y)  
+ {\textstyle\frac{1}{18}} \beta_b \int 
{\rm d}^4x \sqrt{g} \, R^2(x) \, , \label{r1}
\end{eqnarray}
where, since
$\delta_\sigma (\sqrt g R^2) = 12 \sqrt g R \nabla^2 \sigma$, the last
term cancels the $\nabla^2 R$ term in ${\cal G}$.

Just as in two dimensions we may obtain correlation functions
involving the energy momentum tensor by functionally differentiating
the action and then restricting to flat space. However, as shown in \cite{EO},
this does not give rise to conformally invariant results for the flat space 
correlation functions of the energy momentum tensor. 
This appears to be connected with the fact that for the operator
$\Delta_4$ $d=4$ is a critical dimension. For the second order operator
$\Delta_2$ defined in (\ref{2dop}) the critical dimension is $d=2$,
 as is revealed
in (\ref{0form}). From (\ref{0form}) and (\ref{0form2}) we may find a finite
limit for the variation,
\begin{equation}
 \lim_{d\to 2} \oG{}'{}_{\!\!2,\alpha\beta}(x,y;z) 
= \frac{1}{2(2\pi)^2} \, \Big ( Z_\alpha Z_\beta
- \half \de_{\alpha\beta} Z^2\Big ) \, .
\label{G'2}
\end{equation}
Using a complex basis and the explicit form for $Z_\alpha$ in (\ref{Zdef})
we then get
\begin{eqnarray}
\oG{}'{}_{\!\!2, zz}(x_1,x_2;x_3)\Big |_{d\to 2} &=& 
\frac{1}{2(4\pi)^2} \, \frac{(z_1-z_2)^2}
{(z_1 - z_3)^2 (z_2 - z_3)^2} \nonumber \\
&=& \frac{1}{2(4\pi)^2} \bigg ( \frac{1}{(z_1 - z_3)^2} + 
\frac{1}{(z_2 - z_3)^2} - \frac{2}{(z_1 - z_3) (z_2 - z_3)} \bigg ) \, .
\label{varG2}
\end{eqnarray}
It is interesting to compare this conformally covariant limit with the result
shown in (\ref{varG}) which was obtained directly in two dimensions and is
not in accord with conformal invariance (although (\ref{varG}) 
and (\ref{varG2})
agree on further differentiation with respect to both $z_1$ and $z_2$).

A similar picture emerges in relation to the fourth order operators $\Delta_4$,
defined for arbitrary $d$, and the Riegert operator $\Delta^R$ in four
dimensions. The flat space limit of $G^R$ is also logarithmic since from
(\ref{Hdef}) we may take
\begin{equation} \label{H4}
\oG_4(x) = - \frac{1}{16 \pi^2 } \, \ln \mu^2x^2  \, .
\end{equation}
Following (\ref{G'2}) and using (\ref{G4var2}) we may obtain
\begin{equation}
\lim_{d\to 4} \oG{}'{}_{\!\!4,\alpha\beta}(x,y;z) = 
- \frac{1}{3(4\pi^2)^2} \, \Big ( Z_\alpha Z_\beta
- \quar \de_{\alpha\beta} Z^2\Big )Z^2 \, .
\label{G'4}
\end{equation}
However the variation of the Riegert Green function gives
\begin{equation}
\oG{}^{R\prime}{}_{\!\!\!\alpha\beta}(x,y;z)
= G'{}_{\!\!4,\alpha\beta}(x,y;z) \Big |_{d\to 4}
-  {\textstyle{\frac{1}{12}}}
 \lim_{d \rightarrow 4} \Big( (d-4) \, ( X \, \pr_\alpha
\pr_\beta \pr^2 Y + \pr_\alpha \pr_\beta \pr^2 X \, Y) \Big) \, ,
\end{equation}  
with $X,Y$ as in (\ref{Habb},\ref{Hdef}). The difference between taking
the limit $d\to 4$ before or after the variation is then
\begin{eqnarray}
\lefteqn{
\oG{}^{R\prime}{}_{\!\!\!\alpha\beta}(x,y;z) - 
{\oG'{}_{\!\!4,\alpha\beta}(x,y;z)} \Big |_{d\to 4}}
\nonumber\\ & & \!\!\!\!\!\!\!\!\!\!\!\!\!\!{}
= \frac{1}{48 \pi^4} \, \bigg[ \frac{ (y-z)_\alpha (y-z)_\beta }{ (y-z)^6 }
+   \frac{ (x-z)_\alpha (x-z)_\beta }{ (x-z)^6 }
-  \quar \de_{\alpha \beta}
\Big( \frac{1}{(y-z)^4} + \frac{1}{(x-z)^4} \Big) \bigg] \, .
\label{GR'}
\end{eqnarray}
The result for taking the limit $d\to 4$ after the variation given in
(\ref{G'4}) is explicitly conformally invariant but correspondingly the
expression given by (\ref{GR'}) for the variation of the Riegert Green
function shows that the additional terms violate the naively expected
conformal invariance. The additional terms in $G^{R\prime}$ ensure that is
less singular as $x,y\to z$ and they disappear on taking derivatives
both with respect to $x$ and $y$. The results obtained in this paper
therefore show a very strong parellelism between the properties of the second
order operator $\Delta_2$, with critical dimension 2, and the fourth order
operator $\Delta_4$, with critical dimension 4.

\section{Conclusion}

In conclusion we may perhaps reiterate that,
in addition to the results for the fourth order conformal operator discussed
in the previous section, we have shown that the
Green functions of two second order conformal operators lead to conformally
invariant
expressions on flat space when varied with respect to the metric.
For the second order differential operator acting on $k$-forms this expression
is given by (\ref{dgf}), while for the operator on Weyl tensor fields
it is given by (\ref{GCres}).
These results may  be useful for constructing a four-dimensional non-local 
effective action which parallels the two-dimensional Polyakov action 
in at least the following  respects: Like the Polyakov action, this action 
should gene\-rate
the conformal anomaly exactly under Weyl transformations. Moreover, 
when varying this action with respect to the metric,
it should also be possible to  obtain conformal expressions for the 
two and three point functions involving the energy momentum tensor. 

\vspace{1ex}

\section*{Acknowledgements}

\vspace{1ex}

\leftline{I am very grateful to Hugh Osborn for many useful discussions.}

\vspace{3cm}

\renewcommand{\theequation}{E.\arabic{equation}} 
\setcounter{equation}{0}

{\bf Note added.} \hspace{1em} It was pointed out by T.~Branson that 
according to a classification scheme for conformal operators developed in
\cite{Branson3}, the conformal operator on Weyl tensor fields discussed in
section 5 may be written in the form  ${\tilde \Delta}^C = \cal{D}^\wedge{}^* 
\cal{D}$, where $\cal{D}$ is a first order differential operator which 
is conformally covariant in six dimensions. In agreement with these results
it is possible to construct such an operator explicitly.
It is given by
\begin{equation}
(\D \C)_{\mu\si\rho\nu\alpha} =  3\,  \C_{\mu\si[\rho\nu;\alpha]}
- \frac {3}{d-3} \Big ( \nab^\lambda \C_{\lambda\si[\rho\nu}g_{\alpha]\mu}
- \nab^\lambda \C_{\lambda\mu[\rho\nu}g_{\alpha]\si} \Big ) \, ,
\end{equation}
such that  $g^{\mu\alpha}(\D \C)_{\mu\si\rho\nu\alpha} = 0$ and
\begin{eqnarray} \!\!\!\!\!\!\!\!
\delta_\si (\D \C)_{\mu\si\rho\nu\alpha} &=& {\ts\frac{1}{2}} (d-10) \,\si
(\D \C)_{\mu\si\rho\nu\alpha} \nonumber \\
&& {}+ {\ts\frac{3}{2}} (d-6) \bigg \{ \C_{\mu\si[\rho\nu}\pr_{\alpha]}\si
-\frac{1}{d-3}\, \pr^\lambda \si \Big (
\C_{\lambda\si[\rho\nu}g_{\alpha]\mu} - \C_{\lambda\mu[\rho\nu}g_{\alpha]\si}
\Big ) \bigg \} \, .
\end{eqnarray} 
Thus $\D $ is conformally covariant when $d=6$.
Using (E.1) we may write the action ${\tilde S}^C[g,\C]$ 
involving the operator
${\tilde \Delta}^C$ defined in (5.14), letting 
$c= -\frac{1}{4}d, \, e=-\frac{1}{8}d$, in the form
\begin{eqnarray}
{\tilde S}^C[g,\C] &=& \frac{d}{16}  \int \! \d^d x  \sqrt g \,
\bigg \{ \frac{1}{6}\, 
(\D \C)^{\mu\si\rho\nu\alpha}(\D \C)_{\mu\si\rho\nu\alpha}
+ \frac{(d-4)(d-6)}{d(d-3)} \nabla_\mu
\C^{\mu \si \rho \nu} \nabla^\alpha \C_{\alpha \si \rho \nu} \nonumber \\
&& \qquad \qquad\qquad{} -(d-6) \Big ( K_{\mu\alpha} \, \C^{\mu \si \rho \nu}
\C^{\alpha}{}_{ \si \rho \nu}
- {\ts\frac{1}{4}} J\, \C^{\mu \si \rho \nu} \C_{\mu \si \rho \nu} \Big ) 
\bigg \} \, .
\end{eqnarray}
When $d=6$ we may use the identities of section 4 to show that 
$(\D C)_{\mu\si\rho\nu\alpha} = 0$ when acting on the genuine metric dependent
Weyl tensor $C$, so that  ${\tilde S}^C[g,C] = 0$.
When $d=4$ $(\D \C)^{\mu\si\rho\nu\alpha}(\D \C)_{\mu\si\rho\nu\alpha} = 0$
and ${\tilde S}^C[g,\C] = 0$ for any $\C$.

Furthermore it was brought to my attention that the fourth order operator
on scalars constructed by Riegert and also by Paneitz was discussed 
independently in \cite{Tseytlin}. Moreover an early discussion of this operator
in general dimensions as well as of conformal operators acting on vectors and 
on second rank tensors may be found in \cite{Gusynin}.

\vfill\eject
\appendix

\section{Appendix }
\setcounter{equation}{0}
\renewcommand{\theequation}{\thesection.\arabic{equation}}

\subsection{Projection Operator onto the Space of Tensors with Weyl
Symmetry} 

\setcounter{equation}{0} 

The projection operator ${\cal E}^C$ has the explicit form
\begin{eqnarray}
{\cal E}^C{}_{\,\,\mu \sigma \rho \nu,}{}^{\! \alpha \gamma \delta \beta}
& =&
\textstyle{ 1 \over 12} \left( \delta_{\mu}{}^{\! \alpha} 
\delta_{\nu}{}^{\! \beta}
\delta_{\sigma}{}^{\! \gamma} \delta_{\rho}{}^{\! \delta} + 
\delta_{\mu}{}^{\! \delta} \delta_{\sigma}{}^{\! \beta} 
\delta_{\rho}{}^{\! \alpha} \delta_{\nu}{}^{\! \gamma} - \mu
\leftrightarrow \sigma, \nu \leftrightarrow \rho \right) \nonumber\\
& & + \textstyle{ 1 \over 24} \left( 
\delta_{\mu}{}^{\! \alpha} \delta_{\nu}{}^{\! \gamma}
\delta_{\rho}{}^{\! \delta} \delta_{\sigma}{}^{\! \beta} - \mu 
\leftrightarrow \sigma, \nu \leftrightarrow \rho, \alpha
\leftrightarrow \gamma, \beta \leftrightarrow \delta \right)
\nonumber\\
& & - \frac{1}{d-2} \textstyle{\frac{1}{8}}
 \left( g_{\mu \rho} g^{ \alpha \delta} 
\delta_{\sigma}{}^{\! \gamma} \delta_{\nu}{}^{\! \beta} + g_{\mu \rho}
g^{\alpha \delta} \delta_{\sigma}{}^{\! \beta} \delta_{\nu}{}^{\! \gamma} -
\mu \leftrightarrow \sigma, \nu \leftrightarrow \rho, \alpha
\leftrightarrow \gamma , \beta \leftrightarrow \delta \right)
\nonumber\\
& & + \frac{1}{(d-2)(d-1)} \textstyle{\frac{1}{2}} \left( g_{\mu \rho}
g_{\nu \sigma} - g_{\mu \nu} g_{\sigma \rho} \right)
\left( g^{\alpha \delta} g^{\beta \gamma}-  g^{\alpha \beta} 
g^{\gamma \delta} \right) \; .
\end{eqnarray}
This projection operator satisfies the symmetries of the Weyl tensor as given
by (\ref{Weyl}) in both sets of indices.

\subsection{Green Function for the Differential   Operator on $k$-Forms}

Here we calculate the flat space Green function of the operator
defined in (\ref{Delk}).
On flat space $\d\de + \de \d \to - \pr^2$ so from the result in (\ref{Dk})
this operator reduces to
\begin{equation} \label{flatformop}
(\oDel^{(k)} \ome)_{\mu_1 \cdots \mu_k} 
= - (\gamma+ 1) \pr^2 \ome_{\mu_1 \cdots \mu_k} \, + \,
2k \,   \pr_\lambda \pr_{[\mu_1} \ome_{|\lambda |\mu_2  \cdots \mu_k]} \, 
\end{equation}
with $ \gamma = \half(d-2k)$.
Its Fourier transform is
\begin{eqnarray}
P_{\mu_1 \cdots \mu_k, \nu_1 \cdots \nu_k} (p)& =& ( \gamma + 1) 
{\cal E}^A{}_{\!\!\mu_1
\cdots \mu_k, \nu_1 \cdots \nu_k} p^2 \nonumber\\ & & 
{}- 2k\, {\cal E}^A{}_{\!\!\mu_1
\cdots \mu_k, \eps \lambda_2 \cdots \lambda_k } 
\, {\cal E}^A{}_{\!\!\eta \lambda_1
\cdots \lambda_k, \nu_1 \cdots \nu_k} p_\eps p_\eta \, ,
\end{eqnarray}
where ${\cal E}^A$ is the projector on totally antisymmetric tensors
of rank $k$.
The inverse of the Fourier transform is defined by
\begin{equation}
P^{-1}{}_{\!\!\mu_1 \cdots \mu_k, \nu_1 \cdots \nu_k}(p) \,  
P_{\nu_1 \cdots \nu_k, \lambda_1 \cdots \lambda_k}(p) 
= {\cal E}^A{}_{\!\!\mu_1 \cdots \mu_k, \lambda_1 \cdots \lambda_k} \, 
\end{equation}
from which we obtain
\begin{eqnarray}
P^{-1}{}_{\!\!\mu_1 \cdots \mu_k, \nu_1 \cdots \nu_k}(p)
&=& \frac{1}{( \gamma+1)}  {\cal E}^A{}_{\!\!\mu_1
\cdots \mu_k, \nu_1 \cdots \nu_k} \frac{1}{p^2} \nonumber\\ 
& & {} + \frac{2k}{(\gamma+1)(\gamma-1)}
{\cal E}^A{}_{\!\!\mu_1
\cdots \mu_k, \eps \lambda_2 \cdots \lambda_k }\, 
{\cal E}^A{}_{\!\!\eta \lambda_1
\cdots \lambda_k, \nu_1 \cdots \nu_k} \frac{p_\eps p_\eta}{p^4} \, .
\label{inverse}
\end{eqnarray}
When transforming back to $d$ dimensional position space we may use
\begin{equation} \label{fouriert}
\frac{1}{(2\pi)^d} \int \! \d^d p \, e^{-i \, p \cdot x }
 \frac{1}{(p^2)^\alpha} =  \frac{\Gamma (\half d -\alpha)}{4^\alpha
 \pi^{\frac{1}{2}d}\Gamma(\alpha)} \,
\frac{1}{x^{d-2\alpha}} \, , 
\end{equation}
which leads to
\begin{equation} \label{fouriert2} 
\frac{1}{(2\pi)^d} \int \! \d^d p \, e^{-i \, p \cdot x }
\, \frac{ p_{\eps} p_{\eta}}{ p^4} = \,  - \frac{\Gamma(\half d-2)}
{16 \pi^{\frac{1}{2}d}}\, 
\pr_{\eps} \pr_{\eta} \frac{1}{x^{d-4}} \, . 
\end{equation}
With the aid of these relations 
\begin{eqnarray}
\lefteqn{{\widetilde{P^{-1}}}
{}_{\!\!\!\mu_1 \cdots \mu_k, \nu_1 \cdots \nu_k} (x) = 
\frac{\Gamma(\half d)}{ \pi^{\frac{1}{2}d}}  \frac{1}{(d-2k-2)(d-2k+2)} 
\nonumber}\\ && {} 
 \times   \bigg[
 {\cal E}^A{}_{\!\!\mu_1 \cdots \mu_k, \nu_1 \cdots \nu_k} \frac{1}{x^{d-2}} 
 - 2k\, {\cal E}^A{}_{\!\!\mu_1 \cdots \mu_k, \eps \lambda_2 \cdots \lambda_k } 
\, {\cal E}^A{}_{\!\!\eta \lambda_1
\cdots \lambda_k, \nu_1 \cdots \nu_k} \frac{x_\eps x_\eta}{x^{d}} \bigg]
\end{eqnarray}
and finally the Green function for the operator (\ref{flatformop}) becomes
\begin{eqnarray}
\oG^{(k)}{}_{\!\!\mu_1 \cdots \mu_k, \nu_1 \cdots \nu_k}(x)\!& =& \! 
{\widetilde{P^{-1}}}{}_{\!\!\!\mu_1 \cdots \mu_k,\nu_1 \cdots \nu_k} (x)
\nonumber \\
& = & \! \frac{ \Gamma(\half d)}{ \pi^{\frac{1}{2}d} (d-2k-2)(d-2k+2)} \,  
{\cal I}^A{}_{\!\!\mu_1 \cdots \mu_k, \nu_1
\cdots \nu_k} (x)\, \frac{1}{x^{d-2}} \, ,\label{green}
\end{eqnarray}
where ${\cal I}^A$ is the inversion on $k$-forms which is defined
in (\ref{IA}). On functions or $0$-forms this reduces to
\begin{equation} \label{schluss}
\oG^{(0)}(x)= \frac{\Gamma ( \half d)}
{\pi^{\frac{1}{2}d} (d-2)(d+2)}\, \frac{1}{x^{d-2}} \, ,
\end{equation}
which up to a factor $2/(d+2)$ is the standard flat space scalar Green function
as given in (\ref{0form}).

\subsection{Green Function for the Differential Operator on
Weyl Tensor Fields} 

Here we calculate the flat space Green function for 
the operator $\Delta^C$ given by  (\ref{deltac}). 

The flat space reduction of the operator $\Delta^C$ is
\begin{equation}
(\oDel^C \C )_{\mu \si \rho \nu} 
  =   - {\textstyle{\frac{1}{16}}}d\, \pr^2
\C_{\mu \si \rho \nu} 
 + \E^C{}_{\!\!\mu \si \rho \nu, \mu' \si' \rho' \nu'}   
 \pr_{\mu'} \pr_\alpha  \C_{\alpha \si' \rho' \nu'} \, 
\end{equation}
which has the Fourier transform
\begin{equation}
P^C{}_{\!\! \mu \si \rho \nu, \alpha \ga\de\beta} (p) = {\textstyle{1\over
16}} \, d 
\,  p^2 \, 
\E^C{}_{\!\! \mu \si \rho \nu, \alpha \ga\de\beta} - p_\kappa p_\lambda \, 
\E^C{}_{\!\!  \mu \si \rho \nu, \kappa \si' \rho' \nu'}
\E^C{}_{\!\! \alpha \ga \de \beta, \lambda \si' \rho' \nu'} \, .
\end{equation}
The inverse of this Fourier transform is defined by
\begin{equation} 
P^C{}_{\!\! \mu\si\rho\nu, \mu'\si'\rho'\nu'} (p) 
{P^C}^{-1}{}_{\!\!\!\! \mu' \si'
\rho' \nu', \alpha \ga \de \beta} (p) = \E^C{}_{\!\!  \mu \si \rho \nu, \alpha
\ga \de \beta} \, ,
\end{equation}
from which we obtain
\begin{eqnarray}
{P^C}^{-1}{}_{\!\!\!\!\!\! \mu \si \rho  \nu , \alpha \ga \de \beta} (p) &=&
\frac{16}{d} \, \frac{1}{p^2} \, \E^C{}_{\!\!\mu \si \rho  \nu , \alpha \ga \de
\beta}  \nonumber\\ & & 
{}+ \frac{256}{d (d-4)} \, \frac{ p_\eps p_\eta}{p^4} \,
\E^C{}_{\!\!\mu \si \rho
\nu, \eps \varphi \theta \omega}\, \E^C{}_{\!\!\alpha \gamma \delta \beta, \eta
\varphi \theta \omega} \nonumber\\ & & 
{} + \frac{2048}{d(d-4)(d-6)} \, \frac{p_\eps p_\eta p_\kappa
p_\lambda}{p^6} \, \E^C{}_{\!\!\mu \si \rho
\nu, \eps \varphi \eta \omega} \,
\E^C{}_{\!\!\alpha \gamma \delta \beta, \kappa \varphi \lambda \omega} \, .
\end{eqnarray}
To transform this expression back to position space, we use
the equations (\ref{fouriert},\ref{fouriert2}) as well as
\begin{equation} \label{fouriert3}
\frac{1}{(2\pi)^d} \int \! \d^d p \, e^{-ip\cdot x} \, 
\frac{p_\eps p_\eta p_\kappa p_\lambda}{p^6} = 
\frac{\Gamma(\half d-3)}{\pi^{\frac{1}{2}d} 2^7 } \, 
\pr_{\eps} \pr_{\eta} \pr_{\kappa} \pr_{\lambda} \frac{1}{x^{d-6}} \,
\end{equation}
which gives
\begin{eqnarray}
\lefteqn{{\widetilde{{P^C}^{-1}}}{}_{\!\!\!\!\!\!\mu \si \rho  \nu ,
\alpha \ga \de \beta} (x) = \frac{1}{(2\pi)^d}\int \! \d^d p \, e^{-i p\cdot x}
{P^C}^{-1}{}_{\!\!\!\!\!\!\mu \si \rho  \nu , \alpha \ga \de \beta} (p) } 
\nonumber \\
& = & 
\frac{4}{d} \, \frac{\Gamma(\half d-1)}{\pi^{\frac{1}{2}d}} \, \bigg[
\frac{1}{x^{d-2}} \,
\E^C{}_{\!\!\mu \si \rho  \nu , \alpha \ga \de \beta} \nonumber\\ 
& &\hspace{0.2cm} {} + \frac{8}{(d-4)}  \,  \Big(
\frac{1}{x^{d-2}} \, \E^C{}_{\!\!\mu \si \rho  \nu , \alpha \ga \de \beta}
- (d-2) \frac{x_\eps x_\eta}{x^d} \, \E^C{}_{\!\!\mu \si \rho \nu, \eps \varphi
\theta \omega}\, \E^C{}_{\!\!\alpha \ga \de \beta, \eta \varphi \theta \omega}
\Big) \nonumber\\ 
& & \hspace{0.2cm} {} + \frac{16}{(d-4)(d-6)} \,
 \Big( {\textstyle{\frac{3}{2}}}
 \frac{1}{x^{d-2}} \,
\E^C{}_{\!\!\mu \si \rho  \nu , \alpha \ga \de \beta}  
 - 3 (d-2) \frac{x_\eps x_\eta}{x^d} \, \E^C{}_{\!\!\mu \si
\rho \nu, \eps \varphi 
\theta \omega} \, \E^C{}_{\!\!\alpha \ga \de \beta, \eta \varphi \theta \omega} 
\nonumber\\ 
& & \hspace{3cm} {} + d (d-2) \frac{x_\eps x_\eta x_\kappa
x_\lambda}{x^{d+2}} \,  \E^C{}_{\!\!\mu \si \rho
\nu, \eps \varphi 
\eta \omega} \,\E^C{}_{\!\!\alpha \ga \de \beta, \kappa \varphi \lambda \omega}
\Big) \, \bigg] \,  .  
\end{eqnarray}
Now the inversion on the space of tensors with Weyl symmetry 
may be written as
\begin{eqnarray}
\I^C{}_{\!\!\mu \si \rho \nu, \alpha \ga \de \beta} (x) &=&  
\E^C{}_{\!\!\mu \si \rho  \nu , \alpha \ga \de \beta} \, - \, 8 \, 
\E^C{}_{\!\!\mu \si \rho \nu, \eps \varphi 
\theta \omega}\, \E^C{}_{\!\!\alpha \ga \de \beta, \eta \varphi \theta \omega} 
\frac{x_\eps x_\eta}{x^2} \nonumber\\ & & + \, 16 \,  \E^C{}_{\!\!\mu \si \rho
\nu, \eps \varphi  
\eta \omega}\, \E^C{}_{\!\!\alpha \ga \de \beta, \kappa \varphi \lambda \omega}
\frac{x_\eps x_\eta x_\kappa x_\lambda}{x^4}  \, ,
\end{eqnarray}
so that we obtain for the flat space Green function of $\Delta^C$
defined in (\ref{cgdef}) 
\begin{eqnarray}
\oG^C{}_{\!\!\mu \si \rho \nu, \alpha \gamma \delta \beta}  (x) \!&=&\!
{\widetilde{{P^C}^{-1}}}{}_{\!\!\!\!\!\!\mu \si \rho  \nu ,
\alpha \ga \de \beta} (x) \nonumber \\
& = &\! \frac{8 \, \Gamma(\frac{1}{2}d)}{\pi^{\frac{1}{2}d} (d-4)(d-6)} \, 
\frac{1}{x^{d-2}} \, 
\I^C{}_{\!\!\mu \si \rho \nu, \alpha \ga \de \beta} (x) \, ,
\end{eqnarray}
which is in agreement with the form expected  from conformal invariance.

\subsection{Variation of Conformal Actions}

The result for the flat space limit of the metric variation (\ref{Svar}) of 
the $k$-form action (\ref{Sk}) is explicitly
\begin{eqnarray}
\oT^{(k)}{}_{\!\!\! \alpha\beta} &=& 
(\gamma+1) {\textstyle\frac{1}{k!}} \, (\d \A)_{\alpha\mu_1\dots \mu_k}\,
(\d \A)_{\beta \mu_1\dots \mu_k}
- (\gamma -1)  {\textstyle\frac{1}{(k-2)!}} \, 
(\delta \A)_{\alpha\mu_1\dots \mu_{k-2}}\,
(\delta \A)_{\beta\mu_1\dots \mu_{k-2}}  \nonumber \\
&& {}+ 2(\gamma -1) {\textstyle\frac{1}{(k-1)!}}\,
\A_{(\alpha|\mu_1\dots \mu_{k-1}} \, (\d\delta \A)_{\beta)\mu_1\dots \mu_{k-1}}
 \nonumber \\
&& - \half \delta_{\alpha\beta} \Big (
(\gamma + 1) \, (\d \A){\cdot (\d \A)} - (\gamma - 1) \,
(\delta \A){\cdot (\delta \A)} + 2(\gamma - 1) \, \A {\cdot (\d \delta \A)}
\Big ) \nonumber \\
&& {}+ \frac{(\gamma+1)(\gamma -1)}{d-2}\, {\frac{1}{(k-1)!}}\,
\Big ( 2\pr_\mu \pr_{(\alpha} ( \A_{\beta)\mu_1\dots \mu_{k-1}} \,
\A_{\mu\mu_1\dots \mu_{k-1}} ) \nonumber \\
&& \qquad \qquad - \pr^2 ( \A_{\alpha\mu_1\dots \mu_{k-1}}
\A_{\beta\mu_1\dots \mu_{k-1}} ) - \delta_{\alpha\beta}\pr_\mu\pr_\nu
( \A_{\mu\mu_1\dots \mu_{k-1}} \A_{\nu\mu_1\dots \mu_{k-1}} ) \Big )
\nonumber \\
&&{}- \frac{(\gamma+1)(\gamma -1)}{(d-2)(d-1)}(\half d -1 +k)
(\pr_\alpha \pr_\beta - \delta_{\alpha\beta}\pr^2 ) (\A{\cdot \A}) \, .
\end {eqnarray}

The corresponding result for the variation (\ref{SvarC}) reduced to flat
space can be calculated straightforwardly albeit tediously from the
standard form for the dependence of the connection on the metric,
although it is also necessary to take account of the implicit
dependence through the traceless condition $g^{\mu\nu}\C_{\mu\si\rho\nu}=0$.
Without attempting to find the simplest form but leaving the contribution
of each term in the action separate we find
\begin{eqnarray} \label{A20}
\oT^{C}{}_{\!\! \alpha\beta}\!\!\! &=& \!\! {\frac{d}{16}} \Big [
\pr_\alpha \C_{\mu\si\rho\nu} \, \pr_\beta \C_{\mu\si\rho\nu}- 4 \,
\pr^2 \C_{(\alpha|\si\rho\nu}\, \C_{\beta)\si\rho\nu} - \half
\de_{\alpha\beta} \, 
\pr_\gamma \C_{\mu\si\rho\nu} \, \pr_\gamma \C_{\mu\si\rho\nu}
\nonumber \\
&& \qquad -4 \, \pr_\mu (\pr_{(\alpha|} \C_{\mu\si\rho\nu}\, 
\C_{\beta)\si\rho\nu} -  \pr_{(\alpha} \C_{\beta)\si\rho\nu}\,
\C_{\mu\si\rho\nu}) \Big ] \nonumber \\
&&\! {}- 
\pr_\mu \C_{\mu(\alpha|\rho\nu}\, \pr_\lambda \C_{\lambda|\beta)\rho\nu}
-2\, \pr_\mu \C_{\mu\si\rho(\alpha}\, \pr_\lambda \C_{\lambda\si\rho|\beta)}
+2\, \C_{(\alpha|\si\rho\nu}\, \pr_{\beta)}\pr_\mu \C_{\mu\si\rho\nu}
\nonumber \\
&&\! {}- \de_{\alpha\beta} \Big( \half \, \pr_\mu \C_{\mu\si\rho\nu} \,
\pr_\lambda \C_{\lambda\si\rho\nu} + \C_{\lambda\si\rho\nu} \,
\pr_\lambda \pr_\mu \C_{\mu\si\rho\nu} \Big )
\nonumber \\
&&\! {}+ 3 \,\pr_\lambda ( \C_{\lambda(\alpha|\rho\nu}\, \pr_\mu \C_{\mu|\beta)
\rho\nu} ) 
- 2\,\pr_\rho (\C_{\si(\alpha\beta)\nu} \, \pr_\mu \C_{\mu\si\rho\nu})
- {4\over d-2}\, \C_{\alpha\si\beta\nu}\pr_\lambda\pr_\mu\C_{\mu\si\lambda\nu}
\nonumber \\
&&\! {} - \frac{d}{4(d-2)} \Big [ 2\, \pr_\mu \pr_{(\alpha} (
\C_{\beta)\si\rho\nu}\, \C_{\mu\si\rho\nu} ) - \pr^2 
( \C_{(\alpha|\si\rho\nu}\, \C_{\beta)\si\rho\nu} ) - \de_{\alpha\beta}\,
\pr_\mu \pr_\lambda ( \C_{\mu\si\rho\nu}\, \C_{\lambda\si\rho\nu} ) \Big ]
\nonumber \\
&&\! {} - \frac{d(d-6)(d+2)}{64(d-2)(d-1)}\, (\pr_\alpha \pr_\beta -
\de_{\alpha\beta}\, \pr^2) ( \C_{\mu\si\rho\nu} \C_{\mu\si\rho\nu} ) \, . 
\end {eqnarray}

\subsection{Variation of the Weyl Symmetry Green Function}

The tensor $P_{\mu \kappa \lambda \nu, \si \eps \eta 
\rho,\alpha \be}(Z)$ in (\ref{GCres}) has the form
\begin{eqnarray}
\lefteqn{P_{\mu \kappa \lambda \nu, \si \eps \eta 
\rho,\alpha \be}(Z)}   \nonumber\\
&= & \; \, A \, \E^C{}_{\!\!\mu\kappa\lambda\nu,\alpha'\tau\chi\omega}
\E^C{}_{\!\!\si\eps\eta\rho,\beta'\tau\chi\omega}
\E^T{}_{\!\!\alpha'\beta',\alpha\beta}\, {1\over (Z^2)^{\hh d - 2}} \nonumber\\
& &{}  + B\, \E^C{}_{\!\!\mu\kappa\lambda\nu,\si\eps\eta\rho}
\Big ( {Z_\alpha Z_\beta \over Z^2}
- {1\over d} \de_{\alpha\beta} \Big ) {1\over (Z^2)^{\hh d - 2}} \nonumber\\
& &{}  + C \,\E^C{}_{\!\!\mu\kappa\lambda\nu,\alpha'\theta\chi\omega}
\E^C{}_{\!\!\si\eps\eta\rho,\beta'\phi\chi\omega}
\E^T{}_{\!\!\alpha'\beta',\alpha\beta}\, {Z_\theta Z_\phi\over
(Z^2)^{\hh d - 1}} \nonumber\\
& &{}  + D\, \E^C{}_{\!\!\mu\kappa\lambda\nu,\alpha'\chi\omega\theta}
\E^C{}_{\!\!\si\eps\eta\rho,\beta'\chi\omega\phi}
\E^T{}_{\!\!\alpha'\beta',\alpha\beta}\, {Z_\theta Z_\phi\over
(Z^2)^{\hh d - 1}} \nonumber\\
& &{}  + E\, \big ( \E^C{}_{\!\!\mu\kappa\lambda\nu,\alpha'\tau\chi\omega}
\E^C{}_{\!\!\si\eps\eta\rho,\theta\tau\chi\omega}\nonumber\\
& &{} \qquad \qquad + \E^C{}_{\!\!\mu\kappa\lambda\nu,\theta\tau\chi\omega}
\E^C{}_{\!\!\si\eps\eta\rho,\alpha'\tau\chi\omega}\big)
\E^T{}_{\!\!\alpha'\beta',\alpha\beta}\, {Z_\theta Z_{\beta'}\over
(Z^2)^{\hh d - 1}} \nonumber\\
& &{} 
+ F\, \big ( \E^C{}_{\!\!\mu\kappa\lambda\nu,\chi \alpha\beta \omega}
\E^C{}_{\!\!\si\eps\eta\rho,\chi\theta\phi\omega} +
\E^C{}_{\!\!\mu\kappa\lambda\nu,\chi\theta\phi\omega}
\E^C{}_{\!\!\si\eps\eta\rho,\chi \alpha\beta \omega} \big )
{Z_\theta Z_\phi\over (Z^2)^{\hh d - 1}}
 \; , \label{a22}
\end{eqnarray}
with $\E^C$ defined in appendix A.1 and with
$\E^T{}_{\!\! \eps \eta,\alpha\beta} = \half ( \de_{\eps\alpha}
\de_{\eta\beta} + \de_{\eps\beta} \de_{\eta\alpha}) - \frac{1}{d}\de_{\eps\eta}
\de_{\alpha\beta}$. The coefficients are then given by
\begin{eqnarray}
A &=& \frac{d^2}{4 (d-1)(d-2)} (d-16) \nonumber\\
B &=& -\frac{d^2}{64 (d-1)(d-2)} (d^3 - 22 d^2 +76 d - 40) \nonumber\\
C &=& - \frac{d}{ (d-1)(d-2)} (d^3- 2 d^2 +2d -16) \nonumber\\
D &=& \frac{d}{ (d-1)(d-2)} (d^3-8d^2+20d-28) \nonumber\\
E &=& \frac{d}{8 (d-1)(d-2)} (d^4-7 d^3 +24d^2 -36d +48) \nonumber\\
F &=& \frac{d}{2 (d-1)(d-2)} (d^3-8d^2+18d +4) \, .
\end{eqnarray}
These coefficients satisfy two linear relations - derived in \cite{EO} -
which follow from  conservation,
\begin{eqnarray}
\!\!\!\!\!\!\!
{\cal T}_1 \equiv & \frac{1}{2}(d-4)(d+4) A -8 B - \frac{1}{2} (d+2) C
+ \frac{1}{4} (d-4) D    - 2 E - \frac{3}{2} d F  & =0 \, , 
\\ \!\!\!\!\!\!\!
{\cal T}_2 \equiv &
-(d-4)(d+4) A  + 16 B + d C + { 1 \over 4} d (d-4) D    
 - { 1 \over 2 } d
(d-10) F  & = 0  \, \mbox{.}
\end{eqnarray}
This provides a direct check on the conservation of 
${\topcirc{{G}}}{}^{C\prime}$.

\end{document}